\documentclass[conference]{IEEEtran}
\IEEEoverridecommandlockouts
\usepackage{ragged2e}
\usepackage{balance} 
\usepackage{cite}
\usepackage{amsmath,amssymb,amsfonts}
\usepackage{algpseudocode}
\usepackage[lined,linesnumbered,boxed,commentsnumbered, ruled,vlined]{algorithm2e}

\usepackage{etoolbox}
\makeatletter
\patchcmd\algocf@Vline{\vrule}{\vrule \kern-0.4pt}{}{}
\patchcmd\algocf@Vsline{\vrule}{\vrule \kern-0.4pt}{}{}
\makeatother
\usepackage{graphicx}
\usepackage{textcomp}
\usepackage{color}
\usepackage[table,xcdraw]{xcolor}

\usepackage{subfigure}
\usepackage{multirow}
\usepackage{url}
\usepackage{xspace}
\usepackage{mathtools}
\usepackage{enumerate}
\usepackage{amsthm}
\usepackage{float}
\usepackage{caption}
\usepackage{cite}

\newtheorem{theorem}{\textbf{Theorem}}
\newtheorem{definition}{\textbf{Definition}}
\newtheorem{lemma}{\textbf{Lemma}}

\newcommand{\wrt}{\emph{w.r.t.}\xspace}
\newcommand{\ie}{\emph{i.e.,}\xspace}
\newcommand{\eg}{\emph{e.g.,}\xspace}

\newcommand{\eat}[1]{}

\graphicspath{{Figures/}}

\newcommand\colorR[1]{\textcolor{black}{#1}}

\let\svthefootnote\thefootnote
\newcommand\freefootnote[1]{%
  \let\thefootnote\relax%
  \footnotetext{#1}%
  \let\thefootnote\svthefootnote%
}

\def\BibTeX{{\rm B\kern-.05em{\sc i\kern-.025em b}\kern-.08em
    T\kern-.1667em\lower.7ex\hbox{E}\kern-.125emX}}
\begin{document}

\title{High Throughput Shortest Distance Query Processing on Large Dynamic Road Networks}

\author{\IEEEauthorblockN{Xinjie Zhou$^{1,2}$, Mengxuan Zhang$^{3*}$, Lei Li$^{2,1}$, Xiaofang Zhou$^{1,2}$}
	\IEEEauthorblockA{
    {$^1$The Hong Kong University of Science and Technology, Hong Kong SAR, China.}\\
		{$^2$DSA Thrust, The Hong Kong University of Science and Technology (Guangzhou), Guangzhou, China.}\\
  {$^3$School of Computing, The Australian National University, Canberra, Australia.}\\
		{xzhouby@connect.ust.hk}, {Mengxuan.Zhang@anu.edu.au}, {thorli@ust.hk}, {zxf@cse.ust.hk}
	}
}

\maketitle

\begin{abstract}
Shortest path (SP) computation is the building block for many location-based services, and achieving high throughput SP query processing \colorR{with real-time response is crucial for} those services. 
However, \colorR{existing solutions can hardly handle high throughput queries on large dynamic road networks due to either slow query efficiency or poor dynamic adaption.}
In this paper, we \colorR{leverage graph partitioning and propose novel} \textit{Partitioned Shortest Path (PSP)} indexes to address this problem.
Specifically, we first put forward a \textit{cross-boundary strategy} to accelerate the query processing of PSP index and analyze its efficiency upper bound \colorR{theoretically}.
After that, we propose a non-trivial \textit{\underline{P}artitioned \underline{M}ulti-stage \underline{H}ub \underline{L}abeling (PMHL}) that \colorR{subtly aggregates} multiple PSP strategies to achieve fast index maintenance and consecutive query efficiency improvement during index update.
Lastly, to further optimize throughput, we design \textit{tree decomposition-based graph partitioning} and propose \textit{Post-partitioned MHL (PostMHL)} with faster query processing and index update. 
Experiments on real-world road networks show that our methods outperform state-of-the-art baselines in query throughput, yielding up to 2 orders of magnitude improvement.
\end{abstract}

\bstctlcite{IEEEexample:BSTcontrol}

\setlength{\textfloatsep}{-0.8mm} 
\section{Introduction}
\label{sec:Introduction}
\freefootnote{$^*$Mengxuan Zhang is the corresponding author.}

Shortest Path (SP) computation 
on the road networks is a building block for many location-based services (LBSs), such as navigation \cite{bast2016route}, POI recommendation~\cite{POI_li2016point}, logistics, and ride-sharing \cite{cao2015sharek}.
With the increasing utilization of LBSs, three tendencies for SP query emerge: 
1) \underline{\textit{Massive queries}}: 
\textit{Didi} had 462,962 queries per second on average~\cite{ye2018big} in 2018, while Uber faces millions of queries per second. 
Processing SP queries with high \textit{throughput} (\ie the number of queries processed per unit time)
is vital for the real-time response;
2) \underline{\textit{Large-scale networks}}: during 2024 Spring Festival, millions of long-range queries across multiple provinces in China are issued everyday~\cite{SpringFestival}; the interstate highways of the USA handle the highest traffic volume as measured by vehicle-miles traveled \cite{USTransportation}; 
3) \underline{\textit{Traffic dynamicity}}: 
Beijing and New York had about 13.8 and 5.8 edge weight (travel time) updates per second on average \cite{Beijing_zheng2011urban,NY_zhang2012u2sod}. To better support the LBSs, we aim to answer SP queries on large dynamic road networks with high \textit{throughput} in this paper.

\begin{figure}[t]
	\centering
	\includegraphics[width=1\linewidth]{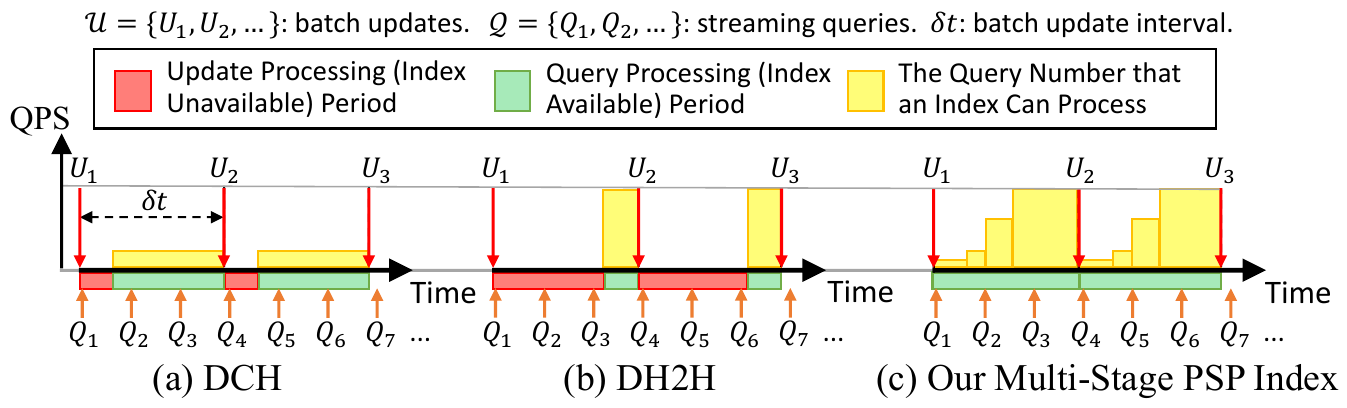}
	\vspace{-0.6cm}
	\caption{Throughput Illustration of Dynamic SP Index \footnotesize{(\colorR{A larger Y-axis value (QPS, Queries per Second) indicates faster query processing; a larger yellow area indicates a higher \textit{QoS-independent}~\cite{qu2006preference} throughput;} length of red and green is index update and query processing time, respectively})}\label{fig:pathFindingSystem}
\end{figure}

The classic SP algorithms such as \textit{Dijkstra's} \cite{dijkstra1959note} and \textit{BiDijkstra} \cite{bidijk_nicholson1966finding} are index-free and can search on the updated network directly but suffer from low efficiency \cite{li2020fast,zhang2020stream}.
The index-based algorithms such as \textit{Contraction Hierarchy (CH)}~\cite{CH_geisberger2008contraction} and \textit{2-Hop Labeling (HL)}~\cite{2hop_cohen2003reachability,HL_abraham2011hub,HHL_abraham2012hierarchical,HopDoubling_jiang2014hop,fu2013label,chang2012exact,TEDI_wei2010tedi,H2H_ouyang2018hierarchy,P2H_chen2021p2h,farhan2023hierarchical} are faster in query with help of shortcuts (CH) and distance labels (HL).
To handle the updates, index maintenance \cite{IncPLL_akiba2014dynamic,DecPLL_d2019fully,IUPLL_qin2017efficient,DPSL_zhang2021efficient,DPSL_zhang2023parallel,DCH_geisberger2012exact,DCH_wei2020architecture,DCH_ouyang2020efficient,DH2H_zhang2021dynamic,BHL_farhan2022batchhl,EPSP_zhang2023universal,DCT_zhou2024scalable} 
are proposed: 
\textit{Dynamic CH (DCH)}~\cite{DCH_ouyang2020efficient} is fast in index update by tracing the shortcut changes, but it is slow in query; \textit{Dynamic Hierarchical 2-Hop labeling (DH2H)}~\cite{DH2H_zhang2021dynamic} is the fastest in query, but is slow in index update due to massive label updates. 
\colorR{
However, the system \textit{throughput} is affected by both query and update efficiency~\cite{TOAIN_luo2018toain,GLAD_he2019efficient}.} 
\colorR{Note that this work follows previous works~\cite{qu2006preference,DCH_ouyang2020efficient,DH2H_zhang2021dynamic,zhang2021experimental,GLAD_he2019efficient} and assumes the update arrives in batch while the system (SP index) should immediately handle the updates before query processing to avoid inaccurate results caused by the outdated index.}
\colorR{As illustrated in Figure \ref{fig:pathFindingSystem}, suppose the edge updates arrive in batches ($U_i$)~\cite{TomTom,INRIX_sharma2017evaluation,AMap},} we start by updating the road network and its SP index, which leads to an index unavailable period (red) for query processing until it is updated to the latest correct version (green).
Although there is a longer time for \textit{DCH} to process queries, its throughput is limited by its low query efficiency; while the throughput of \textit{DH2H} is constrained by its short query processing period. 
\colorR{Worse yet, the \textit{large graph size} and \textit{Quality of Service (QoS)} requirement such as \textit{query response time} (\ie the time taken from the arrival of a query $q$ to the time at which $q$'s answer is computed)~\cite{TOAIN_luo2018toain,GLAD_he2019efficient} could further deteriorate the update time and throughput.}
To this end, \textit{Partitioned Shortest Path (PSP) index} has been investigated for better scalability~\cite{CoreTree_li2020scaling,CRP_delling2017customizable,Gtree_zhong2013g,wang2016effective,li2019time,MCSPs_zhou2024efficient,li2020fastest,yu2020distributed,liu2021efficient} \colorR{and faster index maintenance \cite{EPSP_zhang2023universal}.} 

\textbf{Motivation}. 
Given that the PSP index can exploit the parallelism among partition indexes for fast index maintenance and exhibits good scalability, there comes a question: is it possible to take advantage of this property to achieve our goal?
In other words, 
how about we partition a network into multiple subgraphs with one index for each subgraph, such that we can deal with large networks? Then with the parallelized index update for each subgraph, we can cope with updates quickly and leave more time for query processing. 
Therefore, we aim to propose novel dynamic PSP indexes \colorR{to process high throughput SP queries on large dynamic road networks}.

\textbf{Challenges.} However, it is nontrivial to achieve this goal.

\textit{\underline{Challenge 1}: How to enhance the query efficiency of the PSP index?}
Albeit a fast index update, PSP index sacrifices the query efficiency through distance concatenantion~\cite{EPSP_zhang2023universal}. To this end, we propose a \textit{cross-boundary strategy} to precompute global 2-hop labels and thus accelerate online query processing by eliminating distance concatenation. 
Moreover, we dig into the vertex ordering of PSP index and insightfully reveal the \textit{\colorR{upper bound} of PSP index query efficiency}.

\textit{\underline{Challenge 2}: How to design a PSP index with high query throughput?} High throughput is achieved by fast query and fast updates~\cite{EPSP_zhang2023universal,TOAIN_luo2018toain}.
However, they are hard to balance as a faster query requires a larger index that takes longer time to maintain
\cite{zhang2021experimental,TOAIN_luo2018toain}.
To this end, we propose to improve throughput by taking advantage of the index (un)available time.
\colorR{In particular, we subtly divide the index maintenance interval into multiple stages and} leverage the currently fastest available index or algorithm for query processing for each stage. \colorR{As illustrated in Figure \ref{fig:pathFindingSystem}-(c), the query efficiency is gradually enhanced, thereby improving the query throughput.}
\colorR{To implement this idea,} we first analyze the connection between \textit{DCH} and \textit{DH2H} and put forward an extended \textit{H2H} index called \textit{Multi-stage Hub Labeling (MHL)} that subtly integrates the \textit{CH} and \textit{H2H} index.
Using \textit{MHL} as the underlying SP index, we propose a novel \textit{Partitioned MHL (PMHL)} that incorporates and optimizes multiple PSP strategies to enhance query efficiency across different update stages during index maintenance, thereby achieving high query throughput.


\textit{\underline{Challenge 3}: How to empower \textit{PMHL} with its optimal query efficiency and further accelerate the index update for higher throughput?}
Although the query efficiency is constrained by the aforementioned \colorR{\textit{PSP query upper bound}}, we discover that it can be enhanced with appropriate vertex ordering.
To this end, we put forward \textit{tree decomposition-based graph partitioning (TD-partitioning)} to utilize the higher quality vertex ordering for partitioning. Based on this, we propose a novel \textit{Post-partitioned MHL (PostMHL)} with a redesigned index structure and deliberately optimized PSP strategies to achieve optimal query efficiency (equivalent to \textit{DH2H}) and further expedite index maintenance for higher query throughput.

\textbf{Contributions}. 
We summarize our contributions as follows:
\begin{list}{$\bullet$}{\leftmargin=1em \itemindent=0em}
    \item 
    We propose a novel \textit{cross-boundary strategy} to 
    enhance the query processing and provide an insightful analysis of the \textit{\colorR{upper bound} of PSP index query efficiency};
    \item 
    We propose a novel \textit{PMHL} index and design various optimizations to achieve fast index maintenance and continuously improved query efficiency during index maintenance;
    \item 
    We propose a novel \textit{TD-partitioning} to leverage good vertex ordering for partitioning and propose the \textit{PostMHL} index with both faster query answering and index update than \textit{PMHL} to further improve query throughput;
    \item 
    Extensive experiments on real-world datasets show that our methods have better query throughput than existing solutions, yielding up to 2 orders of magnitude improvement.
\end{list}


\section{Preliminary}
\label{sec:Preliminary}
Let $G=(V, E)$ be a weighted graph where the vertex set $V$ and edge set $E$ denote road intersections and segments. We use $n=|V|$ and $m=|E|$ to denote the number of vertices and edges. Each edge $e(u,v) \in E$ is associated with a positive weight $|e(u,v)|$ for travel time. For each vertex $v\in V$, we denote its neighbor set as $N_G(v)=\{u|(u,v) \in E\}$ and assign it an order $r_G(v)$ indicating its importance in $G$.
A \textit{path} $p_G$ from $s$ to $t$ is a sequence of consecutive vertices $p_G=\left<s=v_0,...,v_j=t\right>=\left<e_1,...,e_{j}\right>$ with length $len(p_G)=\sum_{e\in p_G} |e|$. 
The \textit{shortest path} $sp_G(s,t)$ from $s$ to $t$ is the path with the minimum length.
The \textit{shortest distance query} $q(s,t)$ aims to return the length of $sp_G(s,t)$ (\ie $d_G(s,t)$).
We omit the notation $G$ when the context is clear.
We consider $G$ as an undirected graph, and our techniques can be easily extended to directed graphs.
The dynamicity in this paper refers to edge weight increase or decrease updates. 


\begin{figure*}[t]
	\centering
	\includegraphics[width=1\linewidth]{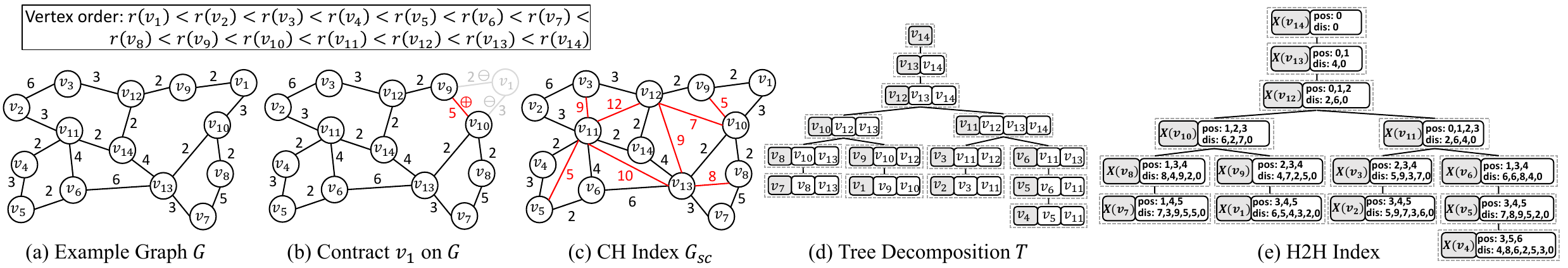}
	\vspace{-0.4cm}
	\caption{Example Road Network $G$, CH Index $G_{sc}$, Tree Decomposition $T$, and H2H Index}\label{fig:exampleGraph}
	\vspace{-0.5cm}
\end{figure*}

\colorR{We next introduce the system model to handle the SP queries. 
We adopt the \textit{batch update arrival} model~\cite{TOAIN_luo2018toain}} since the updates are often processed by batch in most LBSs (\eg TomTom~\cite{TomTom}, INRIX~\cite{INRIX_sharma2017evaluation} \colorR{update every 1 minute while AutoNavi~\cite{AMap} updates every few minutes}) and previous works~\cite{DCH_ouyang2020efficient,DH2H_zhang2021dynamic}. 
\colorR{In particular, we divide time into periods, each of a duration of $\delta t$ seconds. The graph updates $U$ are collected in a batch at the beginning of each period, reflecting the graph changes in the last period.
Following previous works \cite{TOAIN_luo2018toain,GLAD_he2019efficient}, we assume the queries arrive at the system as a Poisson process and are queued for processing.
Besides, following \cite{GLAD_he2019efficient}, when each update batch arrives, we assume the system immediately handles the updates before the query processing to avoid \textit{staleness}~\cite{qu2006preference} (\ie \textit{inaccurate SP computation}). If the update time $t_u\geq\delta t$, the system will have no throughput since it spends all its time on the updates. 
We adopt the \textit{average query response time} $R^*_q$ as a QoS constraint.
Let $t_q, V_q$ be the average and variance of query (processing) time, respectively; and $t_u$ be the average update time; $\lambda_q^*$ be the \textit{maximum average throughput}. We have the following Lemma.}


\vspace{-0.2cm}
\begin{lemma}
\label{lemma:throughput}
\colorR{$\lambda_q^*\leq \min\{\frac{2\cdot(R_q^*-t_q)}{V_q+2\cdot R_q^* \cdot t_q-t_q^2}, \frac{\delta t- t_u}{t_q\cdot\delta t}\}$.}
\end{lemma}
\vspace{-0.3cm}
\begin{proof}
\colorR{The system is an M/G/1 queue. By Pollaczek-Khinchine formula~\cite{cohen2012single}, we have $R_q=\frac{\rho+\lambda_q\mu V_q}{2\cdot(\mu-\lambda_q)}+t_q$, where $\mu=1/t_q$ is the service rate and $\rho=\lambda_q/\mu$. Hence, the first item holds. Besides, fulfilling the update constraint yields $\delta t-\delta t\cdot\lambda_q^*\cdot t_q\geq t_u$, so the second term holds.}
\end{proof}
\vspace{-0.2cm}

Therefore, the \textit{\underline{H}igh \underline{T}hroughput \underline{S}hortest \underline{P}ath query processing (HTSP)} problem aims to maximize the \colorR{\textit{maximum average query throughput} $\lambda_q^*$}.

\section{Existing Solutions}
\label{sec:ExistingSolution}


\subsection{Hierarchy-based Solution}
\label{subsec:Preliminary_CH}
\textit{CH}~\cite{CH_geisberger2008contraction} 
builds a hierarchical \textit{shortcut index} $G_{sc}$ by iteratively contracting vertices in the order of $r(v)$. The contraction involves two steps: 1) for each pair $\forall v_j,v_k\in N(v_i)$, we add the shortcut $sc(v_j,v_k)$ with weight $|sc(v_j,v_k)|=|e(v_i,v_j)|+|e(v_i,v_k)|$ to $G$; 2) we remove $v_i$ and its adjacent edges from $G$ when all pairs of neighbors in $v_i$ are processed. 
We obtain $G_{sc}$ after contracting all vertices. 
The query processing of \textit{CH} is a modified \textit{BiDijkstra}'s search~\cite{bidijk_nicholson1966finding} on 
$G_{sc}$ which only searches the edges/shortcuts from low-order vertices to high-order vertices.
\textit{DCH}~\cite{DCH_ouyang2020efficient}  propose \textit{shortcut-centric} paradigm and \textit{Shortcut Supporting Graph} to identify and propagate the affected shortcuts, resulting in fast index maintenance of the \textit{CH} index. Nevertheless, \textit{DCH} fails to handle high throughput queries due to low query efficiency.


\noindent\textbf{Example 1}. 
\textit{Figure~\ref{fig:exampleGraph}-(a) shows an example network $G$ with vertex order $r(v_1)<...<r(v_{14})$. Figure~\ref{fig:exampleGraph}-(b) illustrates the contraction of vertex $v_1$, adding shortcut $sc(v_9,v_{10})$ to $G$ and eliminating $v_1$ and its adjacent edges. The contraction process is iteratively conducted until all vertices are examined, forming the CH index $G_{sc}$ shown in Figure \ref{fig:exampleGraph}-(c).
}


\subsection{Hop-based Solution}
\label{subsec:Preliminary_H2H}
It assigns each vertex $v\in V$ \colorR{with a vertex set $L(v)\subseteq V$ and a label set $\{(u,L(v,u))|u\in L(v)\}$, where $L(v,u)=d(v,u)$. The labels should fulfill \textit{2-hop cover} property that $\forall s,t\in G$, $d(s,t)=\min_{c\in L(s)\cap L(t)}\{L(s,c)+L(t,c)\}$}.
\textit{Hierarchical 2-Hop Labeling (H2H)}~\cite{H2H_ouyang2018hierarchy} is the state-of-the-art in road networks, which is built based on tree decomposition:

\vspace{-0.1cm}
\begin{definition}[\textbf{Tree Decomposition}]
\label{def:treedecomposition}
The \textit{tree decomposition} $T$ of graph $G=(V,E)$ is a rooted tree in which each tree node $X(v)$ corresponds to a vertex $v\in V$ and is a subset of $V$ ($X(v)\subseteq V$), such that~\cite{H2H_ouyang2018hierarchy}: (1) $\bigcup_{v\in V}X(v)=V$; (2) $\forall (u,w)\in E, \exists X(v),v\in V$ such that $\{u,w\}\subseteq X(v)$; (3) $\forall v\in V$, the set $\{X(u)|v\in X(u)\}$ induces a subtree of $T$.
\end{definition}
\vspace{-0.1cm}

We obtain $T$ through \textit{Minimum Degree Elimination (MDE)}~\cite{MDE_berry2003minimum, MDETD_xu2005tree}. 
In particular, we start by initializing $G$ as $G^0$ and then iteratively contract all vertices. In the $i^{th}$ round ($i\in [1,n]$), we extract minimum-degree vertex $v$ along with its neighboring edges from $G^{i-1}$, generating a tree node $X(v)=\{v\} \cup X(v).N$, where $X(v).N$ are the neighbors of $v$ in $G^{i-1}$. 
We then insert/update the all-pair shortcuts among $X(v).N$ in $G^{i-1}$ to preserve the shortest distances, forming a contracted graph $G^i$. 
The contraction order produces a vertex rank $r$ in ascending order.
After the vertex contraction, $T$ is formed by setting $X(u)$ as the \colorR{parent} of $X(v)$ if $u$ is the lowest-order vertex in $X(v).N$. 
H2H index defines two arrays for each node $X(v)\in T$: \textit{distance array} $X(v).dis$ stores the distances from $v$ to all vertices $u\in X(v).A$, where $X(v).A$ is the ancestor set of $X(v)$; \textit{position array} $X(v).pos$ records the position of neighbor $u\in X(v)$ in $X(v).A$. 

Given a query $q(s,t)$, its shortest distance is calculated as $d(s,t)=\min_{i\in X.pos} \{X(s).dis[i]+X(t).dis[i]\}$ with $X$ denoting the \textit{Lowest Common Ancestor (LCA)}~\cite{LCA_bender2000lca} of $X(s)$ and $X(t)$. 
\textit{DH2H} is the dynamic version of \textit{H2H} which maintains \textit{H2H} index in two phases: \textit{bottom-up shortcut update} and \textit{top-down label update}, which both rely on \textit{star-centric} paradigm to trace and update the affected shortcuts or labels~\cite{DH2H_zhang2021dynamic}.
Nevertheless, its label update phase is time-consuming, making it hard to obtain high throughput.

\noindent\textbf{Example 2}. 
\textit{Figure~\ref{fig:exampleGraph} (d)-(e) presents the tree decomposition $T$ and its H2H index. Given a query $q(v_7,v_4)$, we first identify the LCA of $X(v_7)$ and $X(v_4)$, which is $X(v_{12})$. Then $d(v_7,v_4)=\min_{i\in X(v_{12}).pos}\{X(v_7).dis[i]+X(v_4).dis[i]\}=11$.
}


\subsection{Partition-based Solution}
\label{subsec:Preliminary_PSPIndex}
Graph partitioning can scale up SP index by reducing the construction time \cite{li2019time,liu2021efficient,liu2022fhl}, index size \cite{wang2016effective,li2020fastest,liu2022multi}, and maintenance time~\cite{EPSP_zhang2023universal}. \colorR{We refer to the SP index that leverages graph partitioning as \textit{Partitioned Shortest Path (PSP)} index. As discussed in \cite{EPSP_zhang2023universal}, the PSP index could be classified into three types from the perspective of partition structure: \textit{planar}, \textit{core-periphery}, and \textit{hierarchical} PSP indexes. For example, the classic \textit{CRP}~\cite{CRP_delling2017customizable}, \textit{G-Tree}~\cite{Gtree_zhong2013g}, and \textit{ROAD}~\cite{ROAD_lee2010road} are hierarchical PSP indexes since they organize the partitions hierarchically in multiple levels. The PSP indexes discussed in this paper are all planar PSP indexes that treat all partitions equally in one level.}
In particular, a road network $G$ is divided into multiple subgraphs $\{G_i|1\leq i\leq k\}$ such that $\bigcup_{i\in [1,k]}V(G_i)=V, V(G_i)\cap V(G_j)=\emptyset (\forall i\neq j, i,j\in [1,k])$. 
\colorR{We denote the edges between different partitions on $G$ as $E_{inter}$ while other edges as $E_{intra}$.}
An \textit{overlay graph} $\tilde{G}$ is then built among the boundary vertices of all partitions to preserve the global shortest distances of $G$.
We denote the boundary vertex set of $G_i$ as $B_i$ and $B=\bigcup_{i\in [1,k]}B_i$.
The PSP index $L$ consists of the \textit{partition index} $\{L_i\}$ and the \textit{overlay index} $\tilde{L}$. 
Usually, the PSP index is built by precomputing the distances of all-pair boundary shortcuts \colorR{$E_B(G_i)=\{d(b_{i1},b_{i2})|\forall b_{i1},b_{i2}\in B_i\}$} for each partition $G_i$ by \textit{Dijkstra}'s algorithm. These shortcuts are then inserted into $G_i$ to form an \textit{extended partition} $G'_i$ and are also used to construct $\tilde{G}$ \colorR{($V(\tilde{G})=B,E(\tilde{G})=\bigcup_{i\in[1,k]} E_B(G_i)\cup E_{inter}$)}. $L_i$ and  $\tilde{L}$ are built on $G'_i$ and $\tilde{G}$, respectively.
The above approach is called \underline{\textit{Pre-boundary Strategy}}~\cite{EPSP_zhang2023universal}, whose query processing involves two cases:

\textit{Case 1: Same-Partition}: $\forall s,t\in G_i$, $q(s,t)=d_{L_i}(s,t)$;

\textit{Case 2: Cross-Partition}: $\forall s\in G_i, t\in G_j, i\neq j, q(s,t)=$
\vspace{-0.2cm}
 \begin{footnotesize}
	\begin{equation*}
    \begin{cases}
	d_{\tilde{L}}(s,t) & s,t\in B\\
	\min\limits_{b_{q}\in B_j}\{d_{\tilde{L}}(s, b_{q})+d_{L_j}(b_{q}, t)\} & s\in B, t\notin B\\
	\min\limits_{b_{p}\in B_i}\{d_{L_i}(s, b_{p})+d_{\tilde{L}}(b_{p}, t)\} & s\notin B, t\in B\\
	\min\limits_{b_{p}\in B_i,b_{q}\in B_j}\{d_{L_i}(s, b_{p})+d_{\tilde{L}}(b_{p}, b_{q})+d_{L_j}(b_{q}, t)\} & s\notin B, t\notin B\\
	\end{cases}
	\end{equation*}
 \end{footnotesize}

\begin{figure}[!t]
	\centering
	\includegraphics[width=1\linewidth]{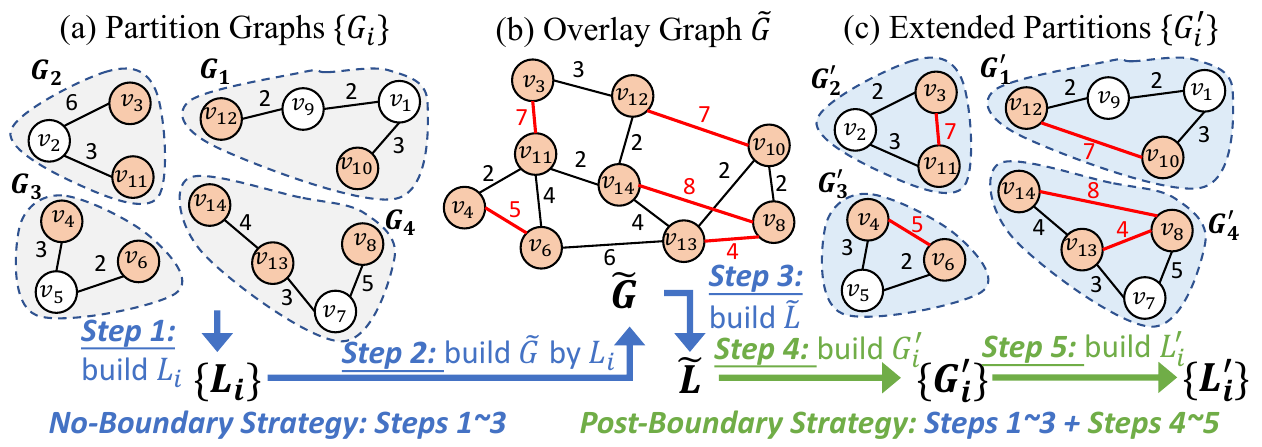}
	\vspace{-0.5cm}
	\caption{\colorR{Illustration of No-boundary and Post-boundary Strategy}}\label{fig:graphPartition}
\end{figure}

\colorR{To enhance the index construction and update of pre-boundary strategy, \cite{EPSP_zhang2023universal} proposes two novel strategies.}

\colorR{\underline{\textit{No-boundary Strategy}}}
\colorR{avoids \textit{Dijkstra}'s searches in boundary shortcut construction and builds the index by 3 \textit{Steps}: 1) build the partition indexes $\{L_i\}$ on $\{G_i\}$; 2) leverage $\{L_i\}$ to compute the boundary shortcuts and construct $\tilde{G}$; 3) construct $\tilde{L}$ on $\tilde{G}$. It subtly converts the boundary shortcut computation from \textit{search-based} to \textit{index-based}, resulting in faster index construction and update.}
Nonetheless, it sacrifices \textit{same-partition} query efficiency due to the concatenation between $\{L_i\}$ and $\tilde{L}$\colorR{: $\forall s,t\in G_i, q(s,t)=\min\{d_{L_i}(s,t),\min_{b_p,b_q\in B_i}\{d_{L_i}(s,b_p)+d_{\tilde{L}}(b_p,b_q)+d_{L_i}(b_q,t)\}\}$.}
\colorR{\textit{N-CH-P}~\cite{EPSP_zhang2023universal} is a update-oriented \textit{no-boundary} PSP index adopting \textit{DCH} as the underlying index}.

\colorR{\underline{\textit{Post-boundary Strategy}}}
\colorR{improves the same-partition querying of no-boundary by adding two more \textit{Steps}: 4) leverage $\tilde{L}$ to compute the all-pair boundary shortcuts and form extended partitions $\{G'_i\}$; 5) build correct partition index $\{L'_i\}$ on $\{G'_i\}$. The post-boundary index $\{\tilde{L},\{L'_i\}\}$ has the same query processing as pre-boundary strategy due to correct partition index.}
\colorR{\textit{P-TD-P}~\cite{EPSP_zhang2023universal} is a \textit{post-boundary PSP index} adopting \textit{DH2H} as the underlying index}.
\colorR{The index update of no-boundary and post-boundary has a similar procedure to their index construction, only replacing the build with index update and performing the update if there is any edge change. More details are introduced in the complete version \cite{full_version}.}
\noindent\textbf{Example 3}. 
\textit{
\colorR{Figure~\ref{fig:graphPartition}-(a) presents the partition graphs $\{G_i\}$,} and (b) shows overlay graph $\tilde{G}$ with pre-computed all-pair shortcuts in red. Extended partitions $\{G'_i\}$ are shown in (c), with orange boundary vertices. \colorR{The index construction procedures are also illustrated at the bottom of Figure~\ref{fig:graphPartition}, where no-boundary strategy involves Steps 1-3 for $\{\tilde{L},\{L_i\}\}$ while post-boundary strategy includes Steps 1-5 for $\{\tilde{L},\{L'_i\}\}$.}}

\noindent\textbf{Remark 1.} 
Although \textit{DCH} can swiftly adapt to dynamic environments, it is relatively slow in query answering, which may necessitate the deployment of hundreds to thousands of servers to address HTSP problem.
\textit{DH2H} offers much higher query efficiency, but its index maintenance is too time-consuming to achieve high query throughput.
\textit{Partition-based solution}~\cite{EPSP_zhang2023universal} provides chances to accelerate the index update of \textit{hop-based solutions} (\ie \textit{DH2H}) by thread parallelism among subgraphs, \colorR{making it promising to leverage the \textit{hop-based solution }for high query throughput in practice}.

\section{Theoretical Analysis of PSP Index}
\label{sec:theoryPSP}
In this section, we first propose a general \textit{Cross-boundary Strategy} for faster PSP query efficiency. Then we provide an insightful analysis of its \colorR{query efficiency upper bound}. 


\subsection{Cross-boundary Strategy}\label{subsec:crossPSP}
Although the PSP index exhibits faster index updates than its non-partitioned counterparts, it sacrifices query efficiency since it has to concatenate the overlay index and partition indexes to obtain the accurate global distance for cross-partition queries~\cite{EPSP_zhang2023universal}.
To improve its query efficiency, we propose the \textit{Cross-boundary Strategy} to build a global non-partitioned cross-boundary index $L^*$ by concatenating the overlay and partition indexes in advance. 
In particular, suppose $\tilde{L}$ and $\{L'_i\}$ have been built by \textit{post-boundary strategy} with 2-hop labeling as the underlying index.
For $\forall v\in B_i$, we \colorR{direct inherit its overlay index}, \ie $L^*(v)=\tilde{L}(v)$ and $L^*(v,c)=\tilde{L}(v,c),\forall c\in \tilde{L}(v)$; otherwise for $\forall v\in G_i\setminus B_i$, $L^*(v)=L'_i(v)\cup \bigcup_{b\in B_i}\tilde{L}(b)$ with the distance label calculated as $L^*(v,c)=\min_{b\in B_i}\{d_{L'_i}(v,b)+d_{\tilde{L}}(b,c)\}, \forall c\in \bigcup_{b\in B_i}\tilde{L}(b)$ \colorR{or $L^*(v,c)=L'_i(v,c), \forall c\in L^*(v)\setminus\bigcup_{b\in B_i}\tilde{L}(b)$}. 
Then we prove the index correctness of $L^*$ in the lemma below:



\begin{lemma}
\label{lemma:crossStrategyCorrect}
$L^*$ satisfies the \textit{2-hop cover property},
 \colorR{\ie $\forall s,t\in G, d(s,t)=\min_{c\in L^*(s)\cap L^*(t)}\{L^*(s,c)+L^*(t,c)\}$}.
\end{lemma}
\vspace{-0.3cm}
\begin{proof}
For $\forall s\in G_i, t\in G_j, i\neq j$, there are three cases based on whether the query endpoints are boundary vertices or not: 
\textit{Case 1}: $s, t\in B$. It holds since $\tilde{L}$ direct inherited overlay index for both $s, t$.
\textit{Case 2}: $s\in B, t\notin B$. 
Suppose $b\in B_j$ lies in $sp(s,t)$, then there must \colorR{contain} a vertex $c\in \tilde{L}(b)\cap \tilde{L}(s)$ in both $sp(s,b)$ and $sp(s,t)$, thus $d(s,t)=L^*(s,c)+L^*(t,c)$.
\textit{Case 3}: $s, t\notin B$. Suppose $b_1\in B_i$, $b_2\in B_j$ lie on $sp(s,t)$, then there must \colorR{contain} a vertex $c\in \tilde{L}(b_1)\cap \tilde{L}(b_2)$ in $sp(b_1,b_2)$ and $sp(s,t)$, thus $d(s,t)=L^*(s,c)+L^*(t,c)$.
\colorR{For $\forall s,t\in G_i$, 
the inherited $L'_i$ naturally satisfies 2-hop cover}.
\end{proof}
\vspace{-0.2cm}

The multi-hop distance concatenations are eliminated and turned to 2-hop concatenation for cross-partition queries with the query complexity reduced by a factor of $O(|B_{max}|^2)$, where $|B_{max}|=\max_{i\in[1,k]}\{|B_i|\}$. 
To update the index, we first maintain $\tilde{L}$ and $\{L'_i\}$. Then for each partition $G_i$, if $\forall b\in B_i, \tilde{L}(b)$ has changed, we update $L^*(v)$ of all non-boundary vertices (\ie $v\in G_i\setminus B_i$). Otherwise, we only check the $L^*(u)$ of non-boundary vertices $u$ with changed $L_i(u)$.

\begin{figure}[t]
	\centering
	\includegraphics[width=1\linewidth]{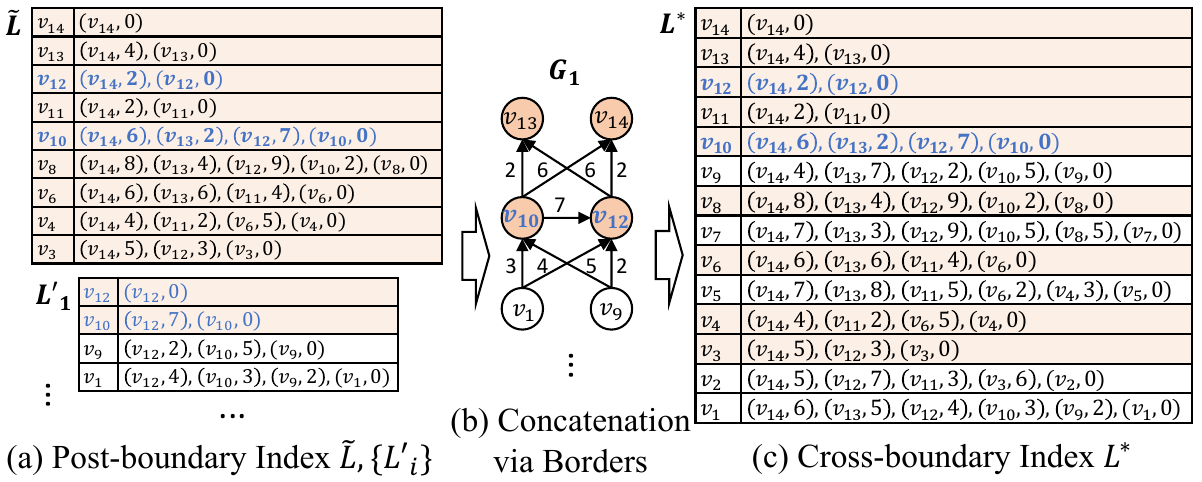}
	\vspace{-0.4cm}
	\caption{\colorR{Illustration of Cross-boundary Strategy}}\label{fig:crossBoundary}
\end{figure}

\noindent\colorR{\noindent\textbf{Example 4}. 
We illustrate the cross-boundary strategy in \textit{Figure~\ref{fig:crossBoundary}. For $v_{10},v_{12}\in B_1$ (blue), we directly inherit their overlay index from $\tilde{L}$. For $v_1,v_9\notin G_1\setminus B_1$, we concatenate $L'_1$ and $\tilde{L}$ via the borders $v_{10}$ and $v_{12}$ to obtain the global distances to $v_{14},v_{13},v_{12},v_{10}$, \eg $L^*(v_1,v_{14})=L'_1(v_1,v_{12})+\tilde{L}(v_{12},v_{14})=6$. 
$L^*$ accelerates cross-partition queries by 2-hop labeling, \eg $q(v_1,v_7)=L^*(v_1,v_{13})+L^*(v_7,v_{13})=8$. 
}}

The \textit{cross-boundary} strategy successfully accelerates query answering by stitching the partitioned indexes and overlay index into a whole one satisfying the 2-hop cover property, which will also be leveraged in Section~\ref{sec:PMHL}.
Nevertheless, we observe that its query answering speed is still inferior to that of \textit{DH2H}. 
To explain this, we next reveal \colorR{the upper bound of PSP index query efficiency theoretically.}

\subsection{\colorR{Upper Bound} of PSP Index Query Efficiency}
\label{subsec:theoryPSP_Curse}
It is well known that vertex order determines the index structure which further affects the SP query efficiency \cite{CH_geisberger2008contraction,H2H_ouyang2018hierarchy,zheng2023reinforcement}.
Hence, we first analyze the order of PSP index under the cross-boundary strategy.
\colorR{Note that to make the discussion meaningful, we assume the 2-hop labeling based on a vertex order $r$ is \textit{canonical labeling}~\cite{HHL_abraham2012hierarchical}, \ie only higher-rank vertex $u$ could be the hub of vertex $v$ and no label is redundant, so it has a minimal label size \wrt order $r$~\cite{HHL_abraham2012hierarchical}.}
We introduce the \underline{\textit{Boundary-first Property}}: for each partition in PSP index, its boundary vertices have a higher order than non-boundary ones. This is because for a cross-partition query $q(s,t)$ ($\forall s\in G_i, t\in G_j)$, $d(s,t)$ is calculated by concatenating distance values $d(s, b_p)+d(b_p,b_q)+d(b_q,t)$ ($\forall b_p\in B_i,b_q\in B_j$). This generally requires non-boundary vertices to store the distances to their boundary vertices, essentially setting a higher vertex order for the boundary vertices as only a higher-order vertex could be the hub of another vertex in \textit{canonical 2-hop labeling} ~\cite{HHL_abraham2012hierarchical}. 

\begin{figure}[t]
	\centering
	\includegraphics[width=0.9\linewidth]{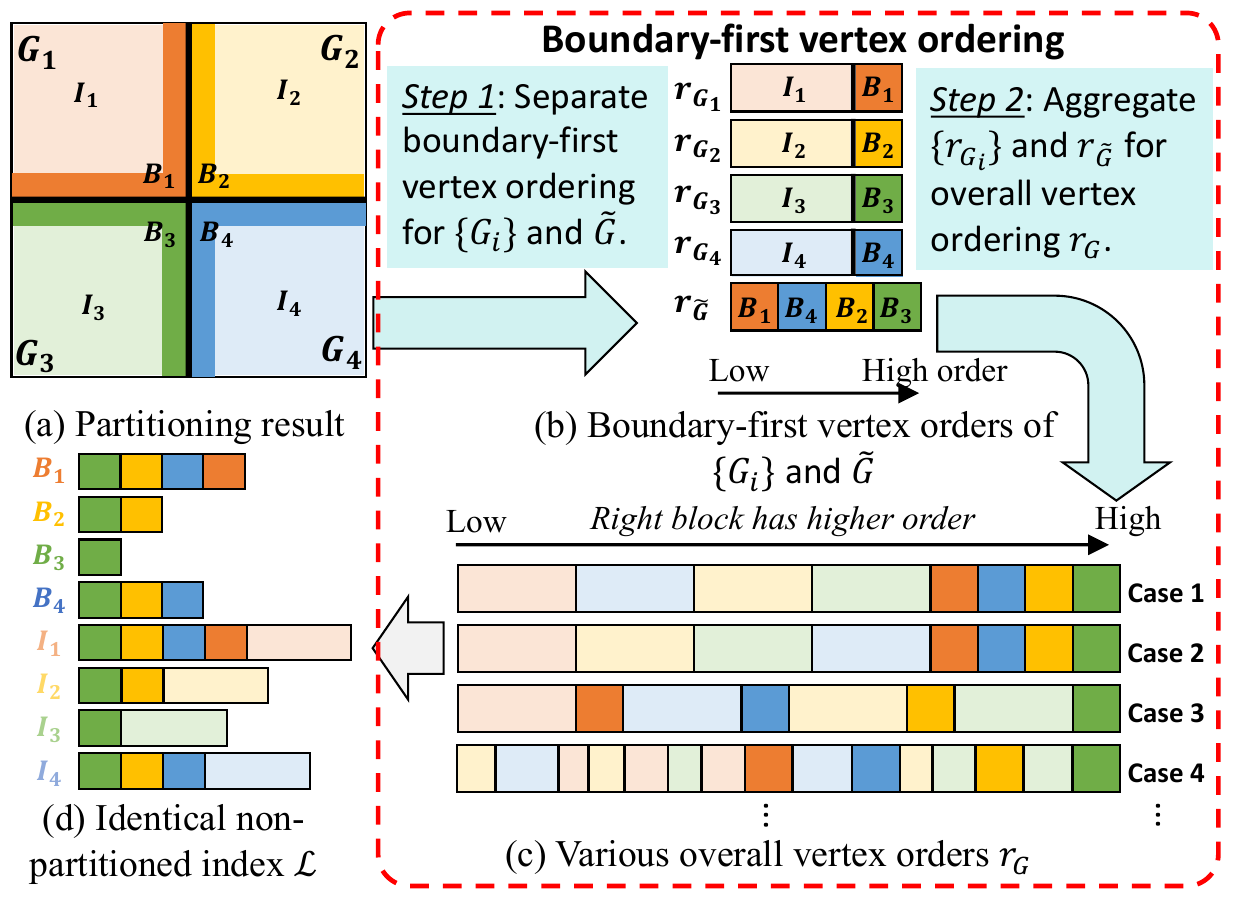}
 \vspace{-0.2cm}
	\caption{Illustration of Boundary-first Vertex Ordering}\label{fig:PSPOrdering}
\end{figure}

Then we discuss how to construct a PSP vertex order satisfying the \textit{boundary-first property}.
As illustrated in Figure~\ref{fig:PSPOrdering}, given a graph partitioning result $\{G_i\}$, we denote the non-boundary vertex set as $I_i$ and boundary vertex set as $B_i$ for each $G_i$. The first step is ordering vertices for all $\{G_i\}$ and $\tilde{G}$ with two conditions: 1) the partition orders $\{r_{G_i}\}$ should satisfy the boundary-first property; 2) the relative order among the boundary vertices in $\{r_{G_i}\}$ should be consistent with overlay order $r_{\tilde{G}}$.
After obtaining the individual vertex orders, the second step is aggregating $r_{\tilde{G}}$ and $\{r_{G_i}\}$ for overall vertex ordering $r_G$ while preserving the relative order in them. 
As shown in Figure~\ref{fig:PSPOrdering}-(c), there are various overall vertex orders satisfying this aggregation requirement. For example, we can set all boundary vertices with higher ranks than non-boundary vertices and randomly interleave the non-boundary vertices (\eg Case 1 and 2).
Nevertheless, the 2-hop labeling based on these different PSP vertex orders leads to identical \textit{canonical labeling} $\mathcal{L}$~\cite{HHL_abraham2012hierarchical} as proved in the following lemma.

\begin{lemma}
\label{lemma:sameCrossIndex}
All vertex orders of a PSP index that satisfy boundary-first property lead to identical \textit{canonical labeling}.
\end{lemma}
\vspace{-0.4cm}
\begin{proof}
Given any boundary-first vertex ordering $r_G$, we discuss whether a vertex $u$ is the hub of $\forall v\in V, r(u)>r(v)$.

\textit{Case 1}: $v\in B$ ($v\in B_i$). When $u\notin B$, $u$ cannot be $v$'s hub if $u\in I_i$ since $r(u)<r(v)$. For the case that $u\in I_j,i\neq j$, $u$ also cannot be $v$'s hub since $sp(u,v)$ must \colorR{contain} a vertex $b\in B_j$ with $r(b)>r(u)>r(v)$, leading to pruned label $\mathcal{L}(v,u)$. Therefore, 
$u$ could be $v$'s hub only if $u\in B$ and $v$'s labels are only determined by $r_{\tilde{G}}$. Since all $r_G$ should preserve the relative order of $r_{\tilde{G}}$, they have identical labels.

\textit{Case 2}: $\forall v\in V\setminus B$. When $u\in I_j,i\neq j$, $u$ cannot be $v$'s hub since $sp(u,v)$ must \colorR{contain} a vertex $b\in B_j$ with $r(b)>r(u)>r(v)$, pruning label $\mathcal{L}(v,u)$. When $u\in B_i$ or $u\in I_i$, $u$ could be $v$'s hub. For $u\in B_j,i\neq j$, $u$ could be $v$'s hub if $r(b)<r(u),\forall b\in B_i$. Hence, $v$'s labels are determined by $r_{G_i}$ and $r_{\tilde{G}}$, which is fulfilled by all $r_G$.
\end{proof}
\vspace{-0.2cm}

\noindent\textbf{Example \colorR{5}}. \textit{Figure~\ref{fig:PSPOrdering}-(d) illustrates the identical non-partitioned index $\mathcal{L}$ based on various $r_G$. For the boundary vertex such as $\forall v\in B_2$, all other higher-order boundary vertices ($u\in B_2\cup B_3$) could be its hub. For the non-boundary vertex such as $\forall v\in I_4$, any higher-order vertex $u\in I_4\cup B_4\cup B_2\cup B_3$ could be its hub since any vertex in $B_3$ and $B_2$ has higher rank than boundary vertex in $B_4$. 
}

\colorR{Lemma~\ref{lemma:sameCrossIndex} reveals that although a PSP index could have various boundary-first orders, the canonical 2-hop labels based on these orders are identical, leading to Theorem~\ref{theorem:partitionToOrder}.}
\vspace{-0.2cm}
\begin{theorem}[\textbf{The \colorR{Upper Bound} of PSP Query Efficiency}]
\label{theorem:partitionToOrder}
    Given a PSP index $L$ and its boundary-first order $r_G$, the optimal query efficiency that $L$ (or \colorR{its cross-boundary index} $L^*$) can achieve is that of \colorR{canonical labeling $\mathcal{L}$ based on $r_G$}.
\end{theorem}
\vspace{-0.3cm}
\begin{proof}
We prove by dividing all scenarios into two cases and indicating that $\mathcal{L}$ is the subset of or equal to $L^*$:
\textit{Case 1}: $\forall v\in B$. $L^*(v)$ and $\mathcal{L}(v)$ have identical index as per Lemma \ref{lemma:sameCrossIndex};
\textit{Case 2}: $\forall v\in V\setminus B$ such as $v\in I_i$. We have $\mathcal{L}(v)\subseteq L^*(v)$ since $\mathcal{L}(v)$ is canonical while $L^*(v)$ may exist redundant labels. For example, suppose $b_1,b_2\in B_i, r(b_1)>r(b_2)$ and $b_1$ is the highest-order vertex in the shortest path between $v$ and $b_2$, then $\mathcal{L}(v,b_2)$ is pruned. By contrast, $b_2$ exists in $L^*(v)$ as per the cross-boundary strategy. Therefore, $\mathcal{L}\subseteq L^*$.
\end{proof}
\vspace{-0.1cm}

This theorem theoretically explains the inferior query efficiency of the existing PSP index, as no graph partitioning method is designed toward a high-quality vertex ordering. Such a pessimistic conclusion seems to restrain any attempt to improve query efficiency through PSP index. But should we really give up? In Section~\ref{sec:PostMHL}, we propose to utilize this theorem reversely by introducing a novel vertex ordering-oriented partitioning to optimize PSP query efficiency.





\section{Partitioned Multi-Stage 2-Hop Labeling}
\label{sec:PMHL}
We propose our first solution \textit{PMHL} in this Section.

\subsection{Integration of DCH and DH2H}
\label{subsec:connectCHandH2H}
As analyzed in Section \ref{sec:ExistingSolution}, there is a trade-off between query and update efficiency for both \textit{DCH} and \textit{DH2H}. Nevertheless, we discover a connection between them that leverages both of their benefits, as shown in the lemma below:

\vspace{-0.2cm}
\begin{lemma}
\label{lemma:CHandH2H}
DH2H can generate equivalent shortcuts required by DCH if they use the same vertex order $r$.
\end{lemma}
\vspace{-0.4cm}
\begin{proof}
As introduced in Section~\ref{sec:Preliminary}, \textit{DCH} and \textit{DH2H} have identical vertex contraction processes (both iteratively contract vertices as per $r$). Therefore, the shortcut generated in the tree decomposition of H2H is equivalent to the shortcut index of CH if they both use the MDE-based vertex ordering~\cite{MDE_berry2003minimum}.
\end{proof}
\vspace{-0.1cm}


Lemma~\ref{lemma:CHandH2H} reveals that the CH construction (resp. update) could be regarded as the first step of H2H construction (resp. update).
Therefore, we could integrate CH into H2H by extending each tree node $X(v),\forall v\in V$ in H2H with an additional \textit{shortcut array} $X(v).sc$ storing distances from $v$ to vertices in $X(v).N$. 
We call such an extended H2H as \textit{\underline{M}ulti-stage \underline{H}ierarchical 2-hop \underline{L}abeling (MHL)} since it provides an intermediate stage to enable the release of CH-based query during H2H index maintenance. 
Moreover, we could leverage \textit{BiDijkstra}~\cite{bidijk_nicholson1966finding} to process queries when both the CH and H2H are unavailable. 
Consequently, \textit{MHL} achieves higher query throughput than the solution solely based on \textit{BiDijkstra}, \textit{DCH} or \textit{DH2H}.
\colorR{We call the scheme of taking full advantage of the fastest available index or algorithm for query processing during index maintenance as \textit{multi-stage scheme}.}

\begin{figure}[t]
	\centering
	\includegraphics[width=1\linewidth]{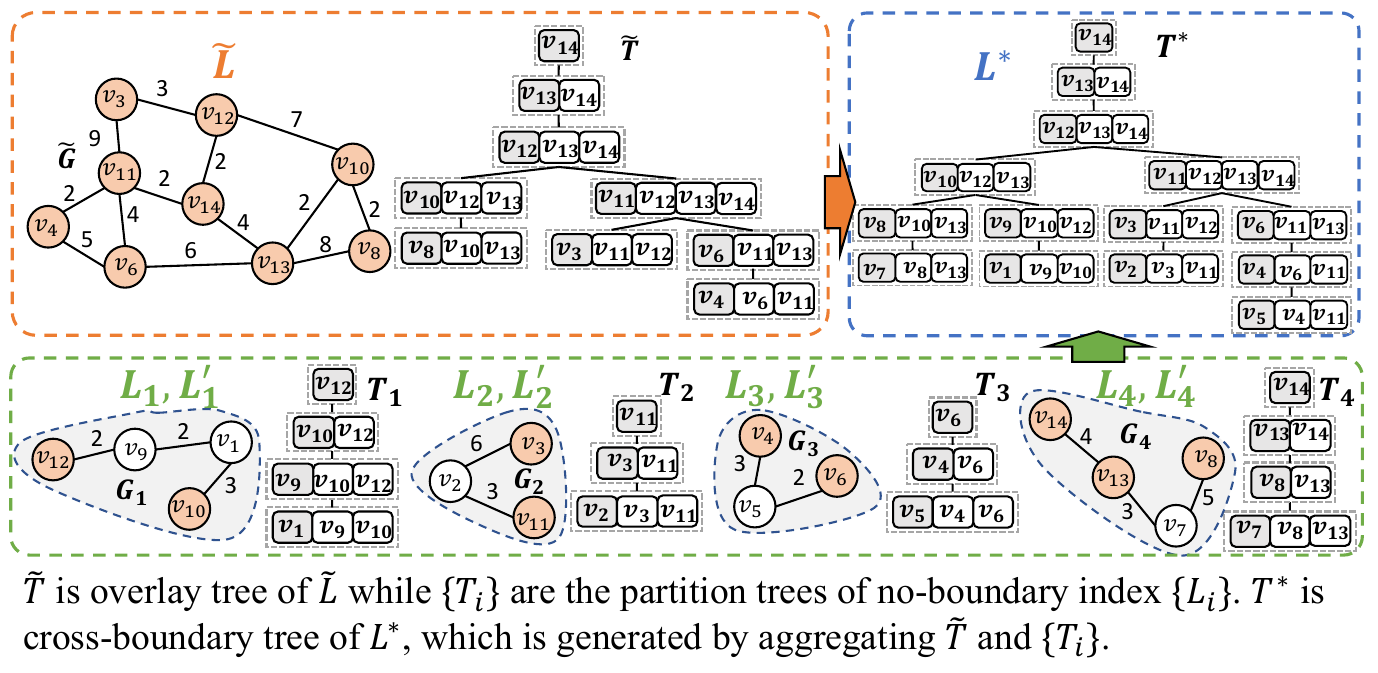}
	\vspace{-0.5cm}
	\caption{Illustration of PMHL Index}\label{fig:PMHLIndex}
\end{figure}

\subsection{Overview of PMHL-based Solution}
\label{subsec:PMHLoverview}
Even with the multi-stage scheme, \textit{MHL} also inherits the laborious maintenance of \textit{DH2H}. As such, the H2H query could be unavailable when the maintenance is time-consuming due to numerous updates or large network size, resulting in low throughput. Therefore, we propose \textit{\underline{P}artitioned \underline{M}ulti-stage \underline{H}ub \underline{L}abeling (PMHL)} which accelerate maintenance through multi-thread parallelization~\cite{EPSP_zhang2023universal}. 
In particular, \textit{PMHL} subtly exploits multiple sets of indexes and different PSP strategies (\textit{no-boundary}, \textit{post-boundary}, and \textit{cross-boundary}) to gradually improve query efficiency while updating the index.
Corresponding to the PSP strategies, \textit{PMHL} includes three types of indexes: \textit{no-boundary index $\{\tilde{L}, \{L_i\}\}$}, \textit{post-boundary index $\{L_i'\}$}, and \textit{cross-boundary index $L^*$}. 

\noindent\textbf{Example \colorR{6}}. \textit{
Figure~\ref{fig:PMHLIndex} illustrates the example of PMHL index, which consists of no-boundary $\{L_i\}$ and post-boundary $\{L'_i\}$ for each $\{G_i|i\in[1,4]\}$, overlay index $\tilde{L}$, and cross-boundary index $L^*$.
The underlying index is MHL and the corresponding tree decompositions are also presented.
}

\begin{figure*}[t]
	\centering
	\includegraphics[width=1\linewidth]{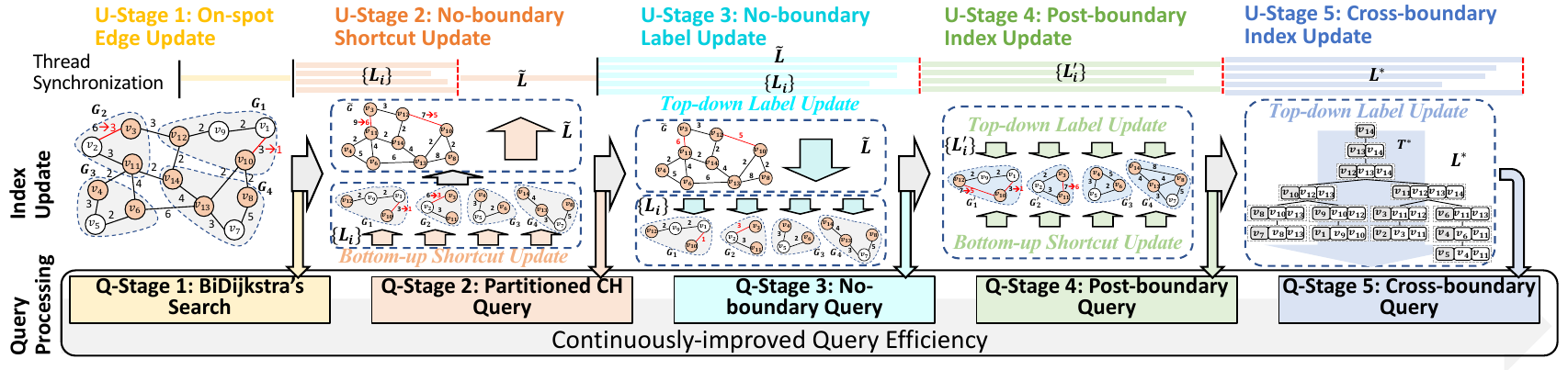}
	\vspace{-0.7cm}
	\caption{Illustration of PMHL-based Solution}\label{fig:PMHLFramework}
	\vspace{-0.4cm}
\end{figure*}

Before diving into the details of \textit{PMHL}, we overview its framework in Figure~\ref{fig:PMHLFramework}. 
There are five stages for update and query. 
After each update stage, it exploits the fastest available index for querying thus continuously enhancing query efficiency.
It is worth noting that since the \textit{DCH} update is the initial step of \textit{DH2H} maintenance (Section~\ref{subsec:connectCHandH2H}), we decouple the no-boundary index update into shortcut update stage (U-Stage 2) and label update stage (U-Stage 3), such that \colorR{\textit{no-boundary Partitioned CH (PCH) query} 
(which adopts the same search mechanism as CH with the only difference that it searches on the union of the shortcut arrays of $\tilde{L}$ and $\{L_i\}$, equivalent to \textit{N-CH-P}~\cite{EPSP_zhang2023universal})} could be released after U-Stage 2.
After U-Stage 5, the cross-boundary index (an H2H-style query processing) is released for the fastest query processing.

\subsection{PMHL Index Construction and Query Processing}
\label{subsec:PMHLIndexing}

To build \textit{PMHL} index, we first leverage \textit{PUNCH}~\cite{PUNCH_delling2011graph} 
to divide $G$ into $k$ subgraphs $\{G_i\}$ and then set the \textit{boundary-first} ordering $r$ with \textit{MDE}~\cite{MDE_berry2003minimum}. Next, we build the \textit{PMHL} index in six steps. \underline{\textit{Step 1}}: Construct index $\{L_i\}$ for each partition $\{G_i\}$ in parallel; \underline{\textit{Step 2}}: Construct the overlay graph $\tilde{G}$ based on the shortcuts among the boundary vertices; \underline{\textit{Step 3}}: Construct index $\tilde{L}$ for $\tilde{G}$; \underline{\textit{Step 4}}: Compute $d_G(b_{i1},b_{i2}),\forall b_{i1},b_{i2}\in B_i$ using $\tilde{L}$ for all partitions $\{G_i\}$ in parallel and insert them to $\{G_i\}$ to get $\{G'_i\}$. \underline{\textit{Step 5}}: Construct $\{L'_i\}$ on all $\{G'_i\}$ in parallel; \underline{\textit{Step 6}}: Construct $L^*$ by aggregating $\tilde{L}$ and $\{L_i\}$.

In particular, Steps 1-3 construct the no-boundary index for \textit{PMHL}, including $\tilde{L}$ and $\{L_i\}$. Steps 4 and 5 focus on building the post-boundary partition indexes $\{L_i'\}$, while Step 6 creates the cross-boundary index $L^*$. We next optimize the index construction and query processing by leveraging different boundary strategy as introduced before.

\subsubsection{\textbf{Optimize Index Construction}}\label{subsubsec:noBoundary}
As introduced in Section~\ref{subsec:Preliminary_PSPIndex}, no-boundary and post-boundary strategies leverage the partition index for the boundary shortcut computation, avoiding the time-consuming \textit{Dijkstra}'s search. 
However, when applying them for \textit{PMHL}, we find that the boundary shortcut computation is still time-consuming due to the low CH query efficiency. To tackle this issue, we optimize these two strategies by directly leveraging the shortcuts formed in the contraction of partition vertices for the overlay graph construction (Step 2). This optimization not only accelerates the index construction and maintenance by eliminating $\sum_{i\in [1,k]}|B_i|\times |B_i|$ queries on $\{L_i\}$, but also reduces the overlay graph size for faster overlay index maintenance. We prove the index correctness after this optimization in Theorem~\ref{theorem:OptimizedOverlay}.

\vspace{-0.1cm}
\begin{theorem}
\label{theorem:OptimizedOverlay}
    The boundary shortcuts generated in the MDE of Step 1 preserve global distances for the overlay graph.
\end{theorem}
\vspace{-0.4cm}
\begin{proof}
For any boundary pair $\forall s,t\in B_i$, we prove $d_G(s,t)$ is preserved in the overlay graph by two cases:
\textit{Case 1}: $sp(s,t)$ does not pass through any non-boundary vertex. This case naturally holds since the contracted non-boundary vertices do not affect the distance;
\textit{Case 2}: $sp(s,t)$ passes through at least one non-boundary vertex. Suppose a non-boundary vertex $u\in B_j$ lays in $sp(s,t)$, we take the concise form of $sp(s,t)$ by extracting only the boundary vertices and $u$ as $sp_c=\left<s=b_0,\dots,b_p,u,b_q,\dots,b_n=t\right>$ ($b_i\in B, 0\le i\le n$), where $b_p$ and $b_q$ are the boundary vertices of $G_j$. The contraction of $u$ and other non-boundary vertices in $G_j$ must lead to shortcut $sc(b_p,b_q)$ in step 1 of optimized no-boundary \textit{MHL}. Therefore, the distance between $b_p$ and $b_q$ is preserved. 
\end{proof}	
\vspace{-0.1cm}

\subsubsection{\textbf{Optimized Query Efficiency}} 
\colorR{As shown in Figure~\ref{fig:PMHLFramework}, the query efficiency of \textit{PMHL} is continuously enhanced in different query stages. \textit{BiDijkstra's} (Q-Stage 1) and \textit{PCH} (Q-Stage 2) are search-based algorithms with relatively slow efficiency. While Q-Stage 3-5 adopt hop-based solutions (based on various PSP strategies) with much faster query efficiency. We next discuss them from the perspective of query type.} 

\colorR{\underline{\textit{Same-Partition Query.}} 
As discussed in Section~\ref{subsec:Preliminary_PSPIndex}, the no-boundary query (Q-Stage 3) has to concatenate $\tilde{L}$ and $\{L_i\}$ for same-partition queries, resulting in slow query processing.
}
However, this query type appears frequently in real-life applications. For instance, it corresponds to city-level queries on province-level road networks. \colorR{To this end, we adopt the post-boundary strategy to fix $\{L_i\}$, obtaining $\{L'_i\}$. As such, the same-partition queries are efficiently answered solely by $L'_i$ in Q-Stage 4 and 5, \ie $\forall s,t\in G_i, q(s,t)=d_{L'_i}(s,t)$}.


\begin{algorithm}[t]
	\caption{\colorR{Cross-boundary PMHL Construction}}
	\label{algo:TDAggregation}
    \scriptsize
	\LinesNumbered
    \colorR{\KwIn{Overlay index $\tilde{L}$, partition indexes $\{L_i\}$}
	\KwOut{Cross-boundary labels $L^{*}=\{X^{*}(v)|v\in V\}$}}
    \colorR{$v\gets$ the highest-order vertex; $A\gets\phi$; //$A$ is ancestor vector\\
    \textsc{TreeAggregation($v,A$)}; \Comment{In a top-down manner} \label{cross:1}\\}
    \SetKwProg{Fn}{Function}{:}{end}
    \Fn{\colorR{\textsc{TreeAggregation}($v,A$)}}
    {
    \colorR{$X^*(v).A\gets A$; //$ch$ is the children set\label{cross:4}\\
    \If{$v\in B$}
    {
        $X^{*}(v).ch\gets \tilde{X}(v).ch\cup X_i(v).ch\setminus B$; //$ch$ is children set\label{cross:4}\\
        $X^*(v).dis\gets \tilde{X}(v).dis$; $X^{*}(v).pos\gets \tilde{X}(v).pos$;\label{cross:5}\\
    }
    \Else{
        $X^{*}(v).ch\gets X_i(v).ch\setminus B$;\label{cross:7}\\
        Compute $X^*(v).dis$ via $X_i(v).N$;\Comment{Compute distance labels}\label{cross:8}\\
    }
    $A.push\_back(v)$;\label{ext:14}\\
    \For{$u\in X^*(v).ch$}
    {
        \colorR{\textsc{TreeAggregation($u,A$)};}\label{cross:10}\\
    }
    $A.pop\_back()$;}\label{ext:17}\\
    }
     
\end{algorithm} 
\colorR{\underline{\textit{Cross-Partition Query.}}
Both no-boundary and post-boundary query processing remains slow as they require distance concatenation for \textit{cross-partition queries} (which can be regarded as cross-province queries).
To solve this, we can utilize the \textit{cross-boundary strategy} discussed in Section~\ref{subsec:crossPSP} to eliminate the distance concatenation for faster query processing for cross-partition queries, \ie $\forall s\in G_i,t\notin G_i, q(s,t)=d_{L^*}(s,t)$.}

However, the naive implementation introduced in Section~\ref{subsec:crossPSP} fails to leverage the tree decomposition for pruning useless label entries, resulting in unsatisfactory query efficiency.
Worse yet, we cannot utilize the \textit{star-centric} paradigm of \textit{DH2H}~\cite{DH2H_zhang2021dynamic} to trace the label changes during index maintenance, leading to slow index update efficiency.
To this end, we propose a \textit{tree decomposition aggregation} method to integrate the overlay tree $\tilde{T}$ and partition trees $\{T_i\}$ for a new cross-boundary tree decomposition $T^*$ and corresponding cross-boundary index $L^*$. 
\colorR{The pseudo-code is presented in Algorithm~\ref{algo:TDAggregation}.}
Starting from the highest-order vertex, \colorR{we compute $L^*$ in a top-down manner.
In particular, for $\forall v\in G_i$, if it is an overlay vertex, we set its children set $X^*(v).ch$ as the union of $\tilde{X}(v).ch$ and $X_i(v).ch\setminus B$ (non-boundary vertices in $X_i(v).ch$); otherwise we set $X^*(v).ch$ as $X_i(v).ch\setminus B$ (Line~\ref{cross:7}).} This integration strategy prioritizes the node relationships of $\tilde{T}$, ensuring that the LCA of the $L^{*}$ aligns with $\tilde{L}$'s LCA for cross-partition queries. 
Besides, $T^*$ also inherits the subordinate relationships of the overlay and partition trees.
\colorR{Therefore, we can directly inherit the distance and position arrays from $\tilde{L}$ for the overlay vertices (Line~\ref{cross:5}).}  For non-boundary vertex $v$, its distance array is computed by using the neighbor set of $v$'s partition index as the vertex separator \colorR{(Line~\ref{cross:10}), \ie $L^*(v,c)=\min_{u\in X_i(v).N}\{|sc(v,u)|+d_{L^*}(u,c)\},\forall c\in X^*(v).A$, where $|sc(v,u)|$ is the shortcut length from $v$ to $u$ on $G_i$}.

\noindent\textbf{Example \colorR{7}}. \textit{
We illustrate the tree aggregation by referring to Figure~\ref{fig:PMHLIndex}. Sourcing from the highest vertex $v_{14}$, we perform the aggregation in a top-down manner. For $v_{14}$, \colorR{it has one child $v_{13}$ according to the union of its child set in $\tilde{T}$ and $T_4$. 
For $v_{10}$, its has two children in $T^*$: $v_8$ from $\tilde{T}$ and $v_9$ from $T_1$.}
}


\vspace{-0.1cm}
\begin{theorem}
\label{theorem:CrossBoundaryCorrect}
$L^*$ based on tree decomposition aggregation satisfies the \textit{2-hop cover property} for cross-partition queries.
\end{theorem}
\vspace{-0.1cm}

The proof of Theorem~\ref{theorem:CrossBoundaryCorrect} is elaborated in our complete version~\cite{full_version} due to limited space.
With the above optimizations on query processing, we can answer the same-partition query with $\{L'_i\}$ and cross-partition query with $L^*$ based on the 2-hop cover property during Q-stage 5 in \textit{PMHL}.

\subsection{PMHL Index Maintenance}
\label{subsec:graphPartitioning}
We next introduce the index update of \textit{PMHL}.

\underline{\textit{U-Stage 1: On-spot Edge Update.}} As shown in Figure~\ref{fig:PMHLFramework}, given a batch of updates, we directly modify the edge weights on $G$, enabling \textit{BiDijkstra's search} for query processing.

\underline{\textit{U-Stage 2: No-boundary Shortcut Update.}} 
During this stage, we update the shortcut arrays of the no-boundary index by referring to the \textit{bottom-up shortcut update mechanism}~\cite{DH2H_zhang2021dynamic}. 
To begin with, we classify the edge updates into two types:
inter-edge updates $U_{inter}$ and intra-edge updates $U_{intra}$. 
If $U_{intra}$ is not empty, we first identify the affected partitions $\{G_i\}$ and maintain the affected $\{L_i\}$ in parallel. Meanwhile, the affected boundary shortcuts are identified and inserted into $U_{inter}$. \colorR{After shortcut updates of $\{L_i\}$, we next update the $\tilde{L}$ based on $U_{inter}$.} The highest affected tree nodes of $\{L_i\}$ and $\tilde{L}$ are recorded for the next stage's update. After this stage, we conduct \textit{Partitioned CH}~\cite{EPSP_zhang2023universal} query by searching the union of shortcut arrays of $\{L_i\}$ and $\tilde{L}$. 

\underline{\textit{U-Stage 3: No-boundary Label Update.}} Given the highest affected tree nodes as input, this stage concurrently maintains the distance arrays of $\tilde{L}$ and $\{L_i\}$ via the \textit{top-down label update mechanism}~\cite{DH2H_zhang2021dynamic}. 
We also store the vertices with updated labels in an affected vertex set $V_A$ for U-Stage 5. After that, \textit{no-boundary query processing} with faster query processing is utilized in \textit{Q-Stage 3}. 

\underline{\textit{U-Stage 4: Post-boundary Index Update.}}
In this stage, we first check whether the all-pair boundary shortcuts of extended partitions $\{G'_i\}$ have been changed by querying over updated $\tilde{L}$. If boundary shortcut updates exist, the corresponding post-boundary indexes $\{L'_i\}$ are concurrently updated using both \textit{bottom-up shortcut} and \textit{top-down label update} mechanisms. Then \textit{post-boundary query processing} with faster same-partition query processing is implemented in \textit{Q-Stage 4}.

\underline{\textit{U-Stage 5: Cross-boundary Index Update.}} We update the cross-boundary index $L^{*}$.
In particular, we first identify the affected partitions $G_A$ based on the affected vertex set $V_A$ generated in U-Stage 3. 
For each affected partition $G_j\in G_A$, we select the highest-order vertex $v\in G_j$ as a representative. Next, we remove the vertices that are descendants of other representative vertices in the cross-boundary tree, leaving only the branch roots of all affected vertices (denoted as $V_R$). Lastly, sourcing from these root vertices in $V_R$, we maintain the cross-boundary index $L^{*}$ in a top-down manner. 
After that, \textit{Q-Stage 5} offers the fastest \textit{cross-boundary query processing}. 

\noindent\colorR{\textbf{Example 8}.} 
\textit{\colorR{The thread synchronization is illustrated at the top of Figure~\ref{fig:PMHLIndex}, where the red dash lines indicate the synchronization points. For U-Stage 2-4, each affected partition (or overlay graph) is allocated with one thread. For U-Stage 5, each root vertices in $V_R$ are allocated with one thread.}}
\section{Post-partitioned Multi-stage Hub Labeling}
\label{sec:PostMHL}
In this section, we further propose a novel \textit{\underline{Post}-partitioned \underline{M}ulti-stage \underline{H}ub \underline{L}abeling (PostMHL)} to enhance throughput.

\begin{algorithm}[t]
	\caption{TD-Partitioning}
	\label{algo:TDPartition}
    \scriptsize
	\LinesNumbered
     \SetKwProg{Fn}{Function}{:}{end}
    \Fn{\textsc{TDPartition}($T,\tau,k_e,\beta_l,\beta_u$)}
    {
        $cN[v]\gets 1,\forall v\in V$; //$cN$ is partition size vector\label{tdp:2}\\
        \For{$X(v)\in T$ in a bottom-up manner}{
            \For{$u\in X(v).ch$}
            {
                $cN[v]\gets cN[v]+cN[u]$;\label{tdp:5}\\
            }
        }
        
        $V_C\gets \phi$; // $V_C$ is a vector storing the root candidates\label{tdp:7}\\
        \For{$v\in V$ in decreasing vertex order}{
            \If{$\beta_l\cdot \frac{|V|}{k_e}\leq cN[v]\leq \beta_u\cdot \frac{|V|}{k_e}$ and $|X(v).N|\leq \tau$}
            {   
                $V_C.push\_back(v)$;\label{tdp:10}\\
            }
        }
        \For{$v\in V_C$}{
            \If{$\forall u\in V_R, u$ is not the ancestor of $v$}
            {
                $V_R\gets V_R\cup v$;\label{tdp:13}
            }
            
        }
        Get the partition result $\tilde{G},\{G_i\}$ based on $V_R$;\label{tdp:14}\\
        \Return $\tilde{G},\{G_i\}$;\\
    }
\end{algorithm}

\subsection{Tree Decomposition-based Graph Partitioning}
\label{subsec:TDPartition}

As discussed in Section~\ref{sec:theoryPSP}, the traditional graph partitioning methods are not designed towards good vertex ordering, thus failing to achieve efficient query processing. 
Conversely, can we obtain a suitable graph partitioning from a satisfactory vertex ordering? 
Observe that tree decomposition $T$ generally produces a good vertex ordering for road networks, so why not leverage it to obtain graph partitioning?
To this end, we design a \textit{Tree Decomposition-based Partitioning (TD-Partitioning)} method to obtain partition results from a good vertex ordering.
The key idea is to choose one root vertex $u$ for each partition $G_i$, setting $u$ and its descendants as the \textit{in-partition vertices} and its neighbor set $X(u).N$ as the boundary vertices. This is because $X(u).N$ can serve as a vertex separator (boundary vertices) for the descendants of $X(u)$ and all other tree nodes. 

We present the pseudo-code of \textit{TD-partitioning} in Algorithm~\ref{algo:TDPartition}. The input includes the tree decomposition $T=\{X(v)|v\in V\}$, bandwidth $\tau$, expected partition number $k_e$, 
partition size imbalance ratio $\beta_l$ and $\beta_u$.
$\tau$ aims to limit the boundary vertex number for all partitions, as it can greatly affect the query efficiency of PSP index~\cite{EPSP_zhang2023universal}. $\beta_l$ and $\beta_u$ constrain the imbalance of partition sizes such that workloads among threads are more balanced.
Firstly, we calculate the descendent vertex number of each vertex $v\in V$ in a top-down manner from the root of $T$ (Lines \ref{tdp:2}-\ref{tdp:5}). 
Then, we calculate the root vertex candidates based on the constraints of bandwidth and partition size bounds (Lines \ref{tdp:7}-\ref{tdp:10}). For instance, if the neighbor set size of vertex $v$, $|X(v).N|$, is no larger than the bandwidth $\tau$, and its partition size falls between  $\beta_l\cdot\frac{|V|}{k_e}$ and $\beta_u\cdot\frac{|V|}{k_e}$, then we add $v$ to the candidate set $V_C$.
After that, we use the \textit{minimum overlay strategy} to determine the root vertex for all partitions, obtaining the root vertex set $V_R$ (Lines 12-\ref{tdp:13}).
This strategy aims to minimize the overlay graph size for faster overall index maintenance efficiency since overlay index maintenance cannot be paralleled.
Lastly, we obtain the partition result by treating the root vertex's tree node and its descendants as the partition vertices and the ancestors of all root vertices as vertices of the overlay graph (Line~\ref{tdp:14}). 
The \textit{TD-partitioning} naturally fulfills the \textit{boundary-first property} as the neighbor set of root vertex has higher vertex orders. Moreover, it allows the PSP index to utilize the vertex ordering generated by \textit{MDE-based tree decomposition}, resulting in faster query efficiency. 

\begin{figure}[t]
	\centering
    \vspace*{-0.3cm}
	\includegraphics[width=1\linewidth]{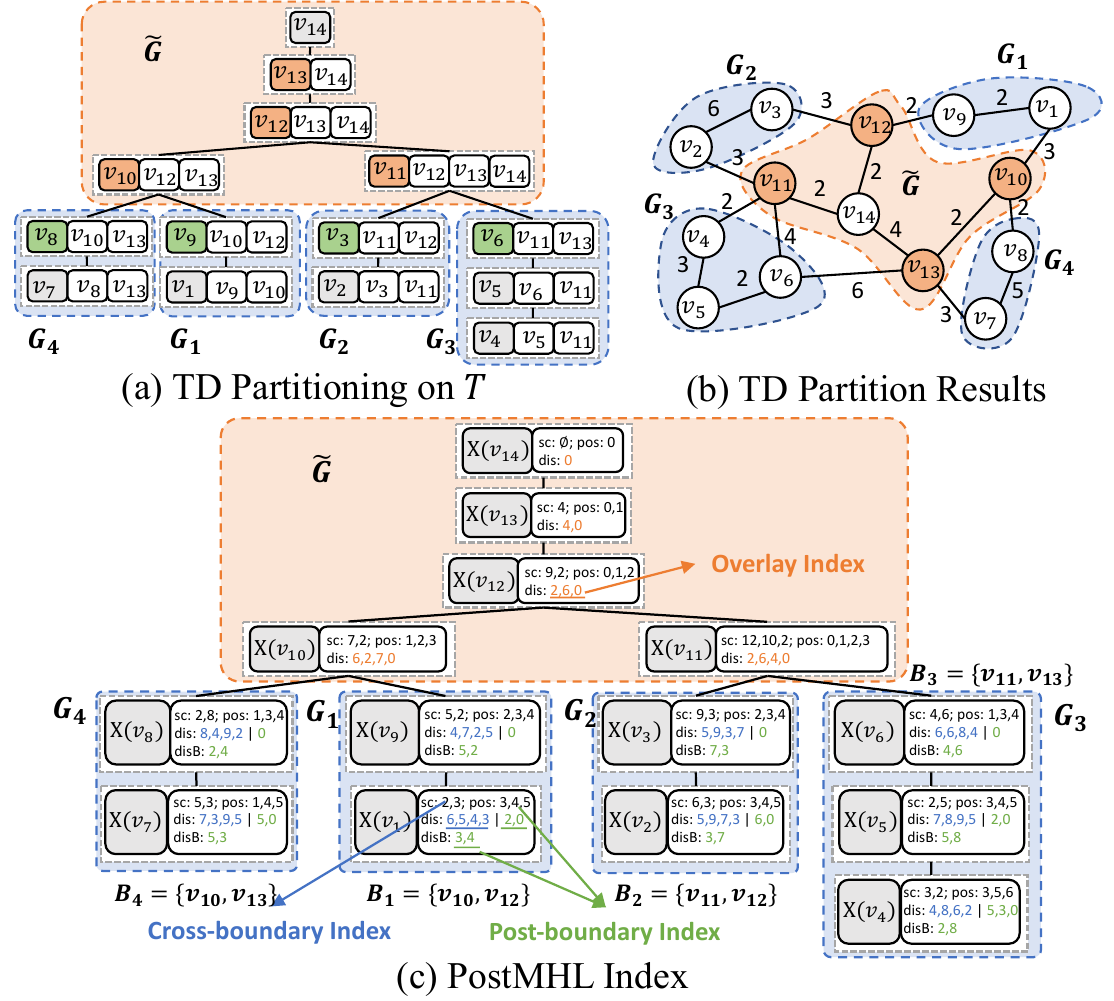}
	\caption{\colorR{TD Partitioning and PostMHL Index}}\label{fig:PostMHLIndex}
\end{figure}



\noindent\textbf{Example \colorR{9}}. 
\textit{Given tree decomposition $T$ and $\tau=2, \beta_l=2, \beta_u=4$, Figure~\ref{fig:PostMHLIndex}-(a) illustrates its TD-partitioning, which consists of four partitions $G_1$ to $G_4$ and an overlay graph $\tilde{G}$.
The root vertices (green) of these partitions are $v_9,v_3,v_6,v_8$, and their boundaries are $\{v_{10},v_{12}\}$, $\{v_{11},v_{12}\}$, $\{v_{11},v_{13}\}$, and $\{v_{10},v_{13}\}$. Figure~\ref{fig:PostMHLIndex}-(b) shows the partition result with orange vertices representing the boundary vertices.
}

\subsection{PostMHL Index}
\label{subsec:PostMHLIndex}

The \textit{TD-partitioning} method empowers the cross-boundary index with better query efficiency. However, updating PSP index is still laborious due to the sequentiality among PSP strategies and multiple tree decomposition.
Fortunately, instead of designing the index from the partition's perspective like \textit{PMHL}, which utilizes no-boundary or post-boundary strategies to guarantee correctness, Theorem \ref{theorem:partitionToOrder} inspires us to re-design it reversely from the nonpartitioned index's perspective and further incorporate the PSP strategies for faster maintenance. 

In particular, \textit{PostMHL} amalgamates \textit{overlay index}, \textit{post-boundary index}, and \textit{cross-boundary index} into one tree decomposition $T$. 
To better illustrate the index component, we classify the ancestors of each tree node into two types: \textit{overlay ancestors} that belong to $\tilde{G}$ and \textit{in-partition ancestors} otherwise. 
Besides, compared with \textit{MHL}, each in-partition node $X(v)$ in \textit{PostMHL} has an additional data structure called \textit{boundary array} $X(v).disB$, which stores the distances from $v\in G_i$ to all boundary vertices $b_p\in B_i$.
As shown in Figure \ref{fig:PostMHLIndex}, the overlay index consists of the distance arrays of all overlay vertices, while post-boundary and cross-boundary indexes exist in the in-partition vertices. The post-boundary index is composed of the distance array entries to the in-partition ancestors and boundary arrays to the boundary vertices, while the cross-boundary index is made up of the distance array entries to the overlay ancestors.
Under this structure, we are safe to drop the no-boundary strategy (as compared with \textit{PMHL}) because the index construction and update of the post-boundary and cross-boundary indexes depend only on the overlay index, as proved in the following Theorem with its proof in our complete version~\cite{full_version} due to limited space.


\begin{figure*}[t]
	\centering
	\includegraphics[width=0.9\linewidth]{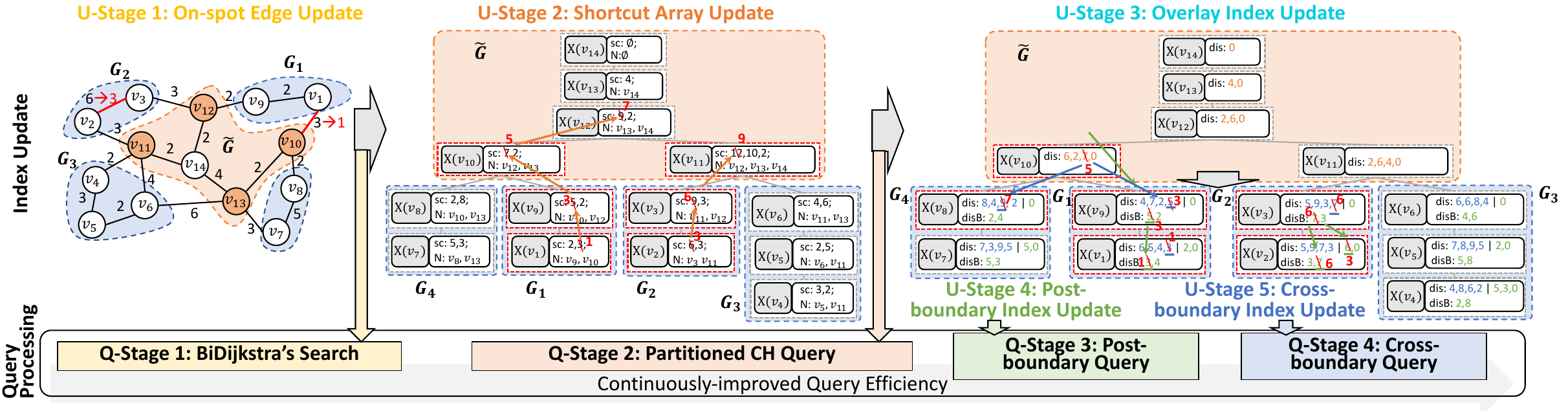}
    \vspace{-0.3cm}	
 \caption{Illustration of PostMHL-based Solution}\label{fig:PostMHLFramework}
    \vspace{-0.4cm}
\end{figure*}
\vspace{-0.2cm}

\begin{theorem}
\label{theorem:overlayIndex}
    The overlay index is sufficient for constructing the post-boundary and cross-boundary index of PostMHL.
\end{theorem}
\vspace{-0.1cm}
\noindent\textbf{Example \colorR{10}}. 
\textit{Figure~\ref{fig:PostMHLIndex}-(c) presents the PostMHL index with the overlay, post-boundary, and cross-boundary indexes in orange, green, and blue. The distance array of $v_1$ is divided into two parts: the entries of overlay ancestors $\{6,5,4,3\}$ for the cross-boundary index and the entries of in-partition ancestors $\{2,0\}$ for the post-boundary index. The post-boundary index also includes the boundary array $X(v_1).disB=\{3,4\}$, which stores the distance from $v_1$ to border vertices $v_{10},v_{12}$. 
The overlay index of $v_{12}$ has a distance array $X(v_{12}).dis=\{2,6,0\}$.}

To build the \textit{PostMHL} index, we first utilize \textit{MDE-based tree decomposition}~\cite{MDETD_xu2005tree} to obtain $T$ and shortcut arrays for all tree nodes.
After that, \textit{TD-partitioning} is employed to obtain the graph partitions and the overlay index is built in a top-down manner like the \textit{H2H} index. 
Then we build the post-boundary index for all partitions in parallel. 
In particular, given the partition $G_i$, we first compute the all-pair distances among the boundary vertices of $G_i$ based on the overlay index and then store the results in a map table $D$ for later reference.
The post-boundary index is constructed in a top-down manner sourcing from the root vertex.
Lastly, the cross-boundary index is constructed in parallel in a top-down manner from each partition's root vertex, similar to the overlay index construction.
The specific pseudo-code is available in our complete version~\cite{full_version}.
We analyze the space complexity and time complexity of \textit{PostMHL} as follows:

\begin{theorem}\label{theorem:PostMHLComplexity}
The space complexity of PostMHL is $O(n\cdot h+n_p\cdot |B_{max}|)$ while the  
index construction complexity is 
$O(n\cdot(w^2+log(n))+n_o\cdot(\tilde{h}\cdot\tilde{w})+\max_{i\in[1,k]}\{|V_i|\cdot(h\cdot w+|B_i|\cdot w)\})$,
where $h$ is the tree height, $|B_{max}|$ is maximal boundary vertex number of all partitions, $n_o$ and $n_p$ are the overlay and in-partition vertex number,
$\tilde{h}$ is the maximum tree height of overlay vertices in $T$, $w$ is treewidth.
The index update complexity of edge decrease in PostMHL is $O(w\cdot(\delta+\Delta\tilde{h}+\frac{\Delta h_p}{k}))$ while edge increase is $O(w\cdot(\delta+(\Delta\tilde{h}+\frac{\Delta h_p}{k})\cdot(w+\epsilon)))$, where $\delta$ is the affected shortcut number, $\Delta \tilde{h}$ and $\Delta h_p$ are the affected tree height in $\tilde{G}$ and $\{G_i\}$, $\epsilon$ is the maximum number of nodes in the subtree involving the affected node.
\end{theorem}

\subsection{Framework of PostMHL}
\label{subsec:PostMHLFramework}
Now, we briefly introduce the framework of \textit{PostMHL} with five update stages and four query stages as illustrated in Figure~\ref{fig:PostMHLFramework}.
The query processing of \textit{PostMHL} is similar to the corresponding Q-Stages in \textit{PMHL}. 
The U-Stage 1 and 2 of \textit{PostMHL} are the same as those of \textit{PMHL}. 
The U-Stage 3 is equivalent to the overlay label update of U-Stage 3 in \textit{PMHL}.
During U-Stage 2 and U-Stage 3, we record the in-partition vertices whose labels are updated in an affected vertex set for each partition to facilitate the post-boundary and cross-boundary index update.
Based on Theorem \ref{theorem:overlayIndex}, the post-boundary and cross-boundary index updates can be conducted in parallel after the overlay index, which further enhances the query throughput with the accelerated index update. Note that since it is more time-consuming to update the cross-boundary index than the post-boundary index due to a larger tree height of the overlay index, the post-boundary index is crucial for improving query throughput as it becomes available earlier.

\noindent\textbf{Example \colorR{11}}. 
\textit{Figure~\ref{fig:PostMHLFramework} shows PostMHL update. U-Stage 1 directly refreshes edges when update batch ($|e(v_2,v_3)|\rightarrow3, |e(v_1,v_{10})|\rightarrow 1$) comes, enabling BiDijkstra's for query. In U-Stage 2, the in-partition shortcuts of affected partitions ($G_1,G_2$) are updated in parallel. 
Meanwhile, the affected overlay shortcuts are recorded and passed to the overlay update. Consequently, all affected shortcuts are refreshed to the correct values in U-Stage 2, enabling the PCH query. 
U-Stage 3 starts with the overlay index update, \eg $X(v_{10}).dis[3]$ is updated from 7 to 5. Then the post-boundary and cross-boundary index updates are performed simultaneously for in-partition vertices. For instance, $v_9$'s post-boundary index, $X(v_9).disB[1]$, is updated from 5 to 3, while $v_8$'s cross-boundary index, $X(v_8).dis[3]$, is updated from 9 to 7.}
 
\noindent\textbf{Remark 2}. 
\textit{PostMHL} further enhances the throughput of \textit{PMHL} by two optimizations: 1) faster index maintenance by eliminating the no-boundary strategy and sharing the same tree decomposition for all indexes; 2) faster query processing by leveraging good vertex ordering obtained from TD-partitioning, achieving equivalent query efficiency with H2H.

\section{Experiments}
\label{sec:Experiments}

\vspace{-0.1cm}
\begin{table}[t]
	\caption{Real-world Datasets}
	\label{table:Dataset}
	\centering
	\scriptsize
    \setlength\tabcolsep{2pt}
    \begin{tabular}{|c|c|r|r|c|c|c|}
    \hline
    \textbf{Name} & \textbf{Dataset}     & $|V|$ & $|E|$  & $k$ &  $k_e$ & $\tau$          \\ \hline
    NY   & New York City$^1$  & 264,346                    & 730,100   &  \colorR{8}     &  \colorR{32}        & 100    \\ \hline
    \colorR{GD}   & \colorR{Guangdong$^2$}  & \colorR{938,957}                    & \colorR{2,452,156}  &  \colorR{8}      &  \colorR{32}       & \colorR{100}    \\ \hline
    FLA   & Florida$^1$      & 1,070,376                  & 2,687,902    &   \colorR{8}    &  \colorR{32}  & 100  \\ \hline
    \colorR{SC}   & \colorR{South China$^2$}   & \colorR{1,326,091}                 & \colorR{3,388,770}   &   \colorR{32}   &  \colorR{64}   & \colorR{200}         \\ \hline
    \colorR{EC}   & \colorR{East China$^2$}  & \colorR{3,008,173}                  & \colorR{7,793,146}   &  \colorR{16}    &  \colorR{32}        & \colorR{200}    \\ \hline
    W   & Western USA$^1$  & 6,262,104                  & 15,119,284   &   16    &   \colorR{32}   & 280        \\ \hline
     CTR   & Central USA$^1$   & 14,081,816                 & 33,866,826   &   32  &  \colorR{128}   & 400         \\ \hline
    \colorR{USA}   & \colorR{Full USA$^1$}   & \colorR{23,947,347}                  & \colorR{57,708,624}   &   \colorR{32}  &  \colorR{128}   & \colorR{400}         \\ \hline

    \end{tabular}
    
  $^1$www.dis.uniroma1.it/challenge9/download.shtml;
  $^2$https://www.navinfo.com/en;
  \vspace*{-0.4cm}
\end{table}

\begin{table}[t]
	\caption{\colorR{Parameters (default values in bold)}}
	\label{table:Parameters}
	\centering
	\scriptsize
    \setlength\tabcolsep{2pt}
    \begin{tabular}{|c|c|}
    \hline
    \colorR{\textbf{Parameters}}    & \colorR{\textbf{Values}}           \\ \hline

    \colorR{Update Volume $|U|$}                & \colorR{500, \textbf{1000}, 3000, 5000}  \\ \hline
    \colorR{Update Interval $\delta t$ (s)}       & \colorR{60, \textbf{120}, 300, 600}   \\ \hline
    \colorR{Query Response Time QoS $R^*_q$ (s)}   & \colorR{0.5, \textbf{1.0}, 1.5, 2} \\ \hline
    
    \end{tabular}
    \vspace*{0.2cm}
\end{table}

\subsection{Experimental Setting}
\label{subsec:Experiments_Setting}
 
\textbf{Datasets.}
We conduct experiments on \colorR{8} real-world road networks from DIMACS~\cite{DIMACSData} \colorR{and NaviInfo~\cite{NavInfoData} (available at \cite{chinaDatasets})}.
Table~\ref{table:Dataset} shows dataset details and the default value of partition number $k$ of \textit{PMHL}, bandwidth $\tau$ and $k_e$ of \textit{PostMHL}.

\textbf{Algorithms.} We compare our approaches (\textit{PMHL} and \textit{PostMHL}) with \colorR{six} baselines: \textit{BiDijkstra}~\cite{bidijk_nicholson1966finding}, \textit{DCH}~\cite{DCH_ouyang2020efficient}, \textit{DH2H}~\cite{DH2H_zhang2021dynamic}, \colorR{\textit{N-CH-P}~\cite{EPSP_zhang2023universal}, \textit{P-TD-P}~\cite{EPSP_zhang2023universal}, and \textit{TOAIN}~\cite{TOAIN_luo2018toain}}. 
\colorR{\textit{DCH} and \textit{DH2H} are state-of-the-art non-partitioned algorithms discussed in Section~\ref{sec:Preliminary}. \textit{N-CH-P} and \textit{P-TD-P} are update-oriented and query-oriented PSP indexes discussed in Section~\ref{subsec:Preliminary_PSPIndex}. \textit{TOAIN} is a throughput-optimizing index for dynamic $k$NN queries and can be used for SP computation by setting $k=1$ for $k$NN. In particular, given a random query $q(s,t)$, we regard $s$ as the query node while $t$ is the nearest object. However, since \textit{TOAIN} is designed for static road networks, we have to adapt it to dynamic networks by refreshing its shortcuts in updated networks.
It is worth noting that to avoid system unresponsiveness, we also leverage index-free \textit{BiDijkstra} to answer the queries for index-based baselines during their index maintenance.}
We set $\beta_l=0.1$, $\beta_u=2$ for \textit{PostMHL} based on our preliminary experiments.
All algorithms are implemented in C++ with full optimization on a server with 4 Xeon Gold 6248 2.6GHz CPUs (total 80 cores / 160 threads) and 1.5TB RAM. 
The default thread number is set as 140 for all PSP algorithms.

\colorR{\textbf{Queries and Updates.}}
\colorR{We generate a random query set $Q$ in a Poisson process with the arrival rate of $\lambda_q$.}
We generate 10 batches of random updates for each dataset. For each update batch $U$, we follow \cite{DCH_ouyang2020efficient,zhang2021experimental} to randomly select 10,000 edges and for each selected edge $e$,  we decrease its edge weight to $0.5\times|e|$ or increase to $2\times|e|$. 
\colorR{The default \textit{update volume} $|U|$ and \textit{update interval} $\delta t$ are shown in bold in Table~\ref{table:Parameters}.}

\colorR{\textbf{Measurements.}}
\colorR{We use \textit{average query response time} $R^*_q$ as QoS and measure the \textit{maximum average query throughput} $\lambda_q^*$. Similar to \cite{TOAIN_luo2018toain}, we run the system on $10\times \delta t$ seconds with a certain query arrival rate $\lambda_q$ and gradually increase $\lambda_q$ and repeat until QoS is violated or the system is overloaded (the updates $U$ cannot be installed in $\delta t$). 
The default $R^*_q$ is shown in Table~\ref{table:Parameters}.} 
We also report the average \textit{index update time} $T_u$, \textit{query time} $T_q$, \textit{indexing time} $T_c$, and \textit{index size} $|L|$.

\begin{figure}[t]
	\centering
	\includegraphics[width=1\linewidth]{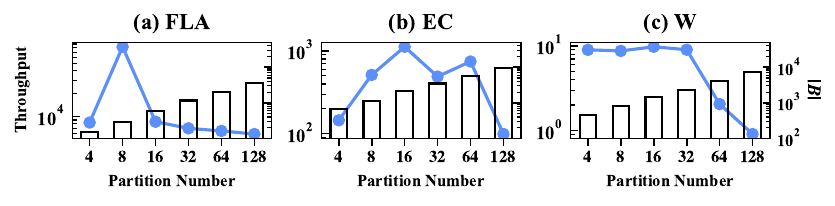}
	\vspace{-0.8cm}
	\caption{\colorR{Effect of Partition Number (Polyline: $\lambda^*_q$; Bar: $|B|$)}}\label{fig:partiNumber}
	\vspace{-0.4cm}
\end{figure}


\begin{figure}[t]
	\centering
    \includegraphics[width=1\linewidth]{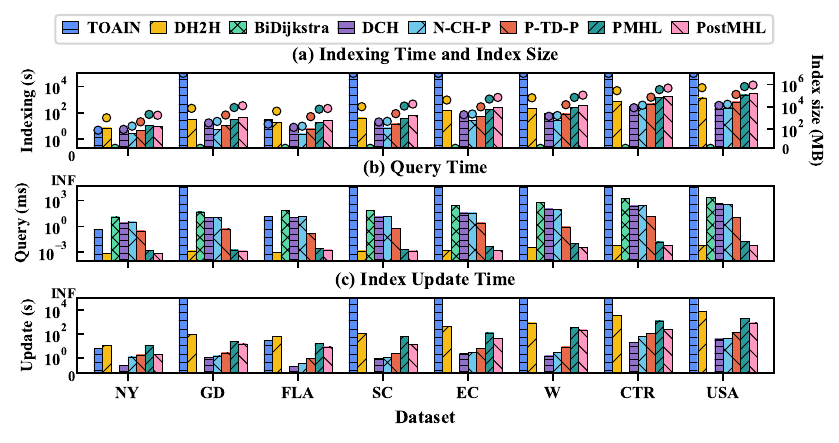}
	\vspace{-0.8cm}
	\caption{\colorR{Index Performance Comparison}}\label{fig:comparePerformance}
\end{figure}

\begin{figure}[t]
	\centering
    \includegraphics[width=1\linewidth]{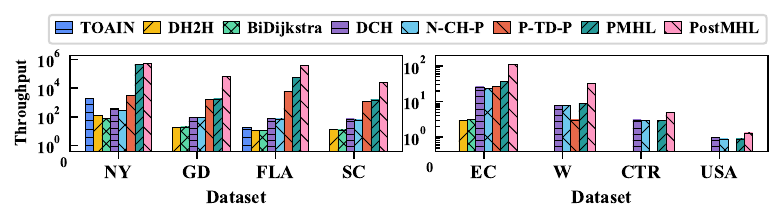}
	\vspace{-0.8cm}
	\caption{\colorR{Throughput Comparison on Different Datasets}}\label{fig:baselineThroughput}
	\vspace{-0.6cm}
\end{figure}


\subsection{Experimental Results}
\label{subsec:Experiments_Results}
\textbf{Exp 1: Effect of Partition Number.}
We first vary partition number $k$ from \colorR{4} to 128 to evaluate its effect on \textit{PMHL}. Figure~\ref{fig:partiNumber} shows the results on \colorR{SC, EC,} and W, where both small and large $k$ result in reduced throughput. \colorR{$k=8$ or $16$} generally performs the best due to the balance between update workload and parallelization. A larger $k$ can better exploit parallelization but increase the boundary number and tree height for the cross-boundary index, deteriorating the update efficiency of overlay and cross-boundary index.
We also test the effect of bandwidth $\tau$ and $k_e$ for \textit{PostMHL}, which is in the Appendix~\cite{full_version} due to limited space. 
The default $k$, $k_e$, and $\tau$ are listed in the last three columns of Table~\ref{table:Dataset}.

\begin{figure}[t]
	\centering
    \includegraphics[width=1\linewidth]{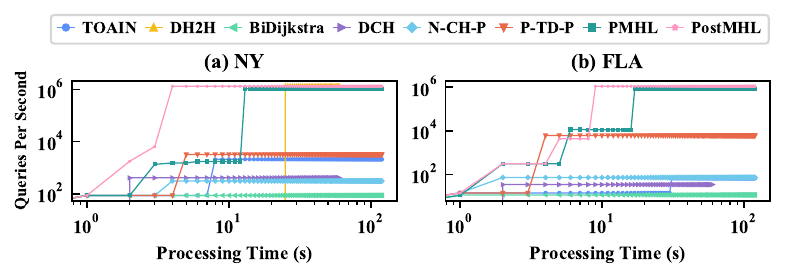}
	\vspace{-0.7cm}
	\caption{Evolution of Queries Per Second}\label{fig:throughputEvolve}
\end{figure}


\begin{figure*}[t]
	\centering
    \includegraphics[width=1\linewidth]{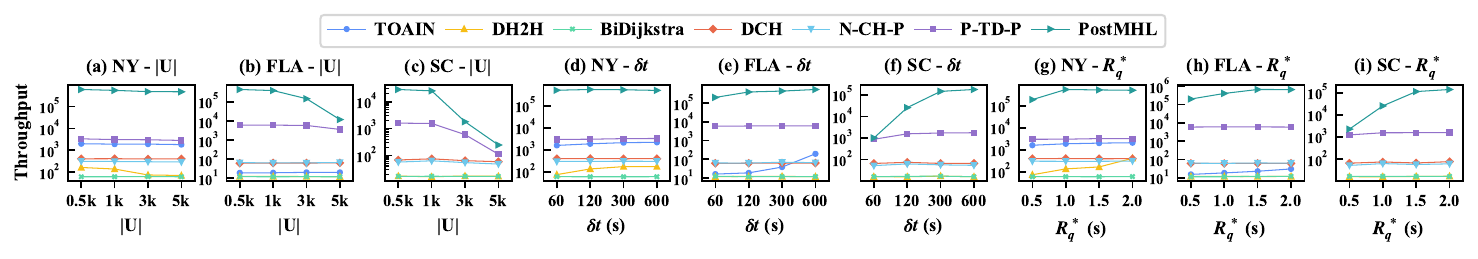}
	\vspace{-0.8cm}
	\caption{\colorR{Throughput Performance When Varying Parameters}}\label{fig:varyParameter}
	\vspace{-0.6cm}
\end{figure*}

\begin{figure}[t]
	\centering
    \includegraphics[width=1\linewidth]{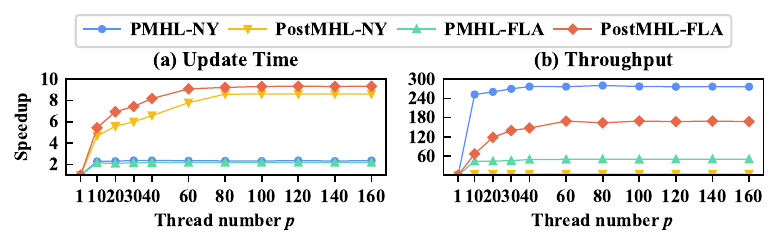}
	\vspace{-0.75cm}
	\caption{\colorR{Speedup When Varying Thread Number}}\label{fig:threadNum}
\end{figure}


\colorR{\textbf{Exp 2: Index Performance Comparison}.}
\colorR{Next, we report the index performance ($t_c, |L|, t_q, t_u$) in Figure~\ref{fig:comparePerformance}. Note that we only report the results of TOAIN on NY and FLA since its index construction on other datasets exceeds 6 hours. 
As shown in (a), although $t_c$ and $|L|$ of \textit{PostMHL} are slightly larger than \textit{DH2H} due to the addition of boundary arrays, it still takes less than an hour to build \textit{PostMHL} index. 
For query time, \textit{PMHL} achieves up to $1000\times$ speedup over \textit{P-TD-P} which only relies on the post-boundary strategy, validating the effectiveness of cross-boundary strategy. Besides, due to the thread parallelization among partition indexes, the index update efficiency of \textit{PostMHL} is about $3.9-15.5\times$ faster than \textit{DH2H}, while the query efficiency is almost the same.}

\colorR{\textbf{Exp 3: Throughput Comparison}.}
\colorR{We next report the performance of throughput $\lambda^*_q$ in Figure~\ref{fig:baselineThroughput}. It is worth noting that we especially set $\delta t=600$ and $R^*_q=5$ for CTR and USA since all algorithms suffer from almost zero throughput in the default setting. Nevertheless, we argue that in real-life applications, concurrent query processing (using multiple threads (workers) for query processing) could be leveraged to enhance the throughput and meet the real-time response for these super-large road networks. However, this provides the same benefits for all algorithms. Hence, we argue that this slacked setting is still a reasonable way for evaluation.}
\colorR{As shown in Figure~\ref{fig:baselineThroughput}, \textit{PostMHL} and \textit{PMHL} outperform all baselines with up to 2 orders of magnitude higher query throughput than the best applicable baselines.
It is worth noting that even with state-of-the-art query efficiency and the help of \textit{BiDijkstra}'s query processing during the index maintenance, the throughput of \textit{DH2H} is still low due to its slow maintenance. \textit{TOAIN} performs well on NY due to its adaptability in trading off query and update. \textit{P-TD-P} and \textit{DCH} are the best baselines in medium-sized or large road networks due to a good tradeoff between the index update and query time. Nevertheless, \textit{PostMHL} always has the best throughput.}



\textbf{Exp \colorR{4}: Evolution of \colorR{Queries Per Second (QPS)}.} 
\colorR{To better understand why our methods have better throughput, we report the evolution of \textit{QPS} (equals to $1/t_q$) during the update interval (120s)} on NY and FLA.
As shown in Figure~\ref{fig:throughputEvolve}, \colorR{as time goes by, \textit{PostMHL} (or \textit{PMHL}) generally has the best query efficiency at any time point because of the three key designs: 1) \textit{multi-stage scheme} to continuously enhance query efficiency; 2) \textit{cross-boundary strategy} for exceptional query efficiency; 3) multi-thread parallelization for fast maintenance.}
Besides, \textit{PostMHL} outperforms \textit{PMHL} in most cases due to the faster maintenance and cross-boundary query.
Since \textit{PostMHL} generally has the best performance, we use it as the representative of our methods in the following experiment.

\textbf{Exp 5: \colorR{Effect of Update Volume, Interval, and Response Time}.}
We vary $|U|$, $\delta t$, \colorR{and $R^*_q$} to evaluate the effectiveness of \textit{PostMHL} under different update scenarios. 
\colorR{As shown in Figure~\ref{fig:varyParameter} (a)-(c), with the increase of update volume $|U|$, the throughput of most SP indexes tends to decrease due to longer index maintenance. Nevertheless, \textit{PostMHL} consistently outperforms all baselines with $2.2-182.9\times$ speed up.
Figure~\ref{fig:varyParameter} (d)-(f) show the throughput of most algorithms (\textit{BiDijkstra}, \textit{DCH}, \textit{N-CH-P}) remains stable with the increase of update interval $\delta t$, demonstrating that they have reached their best performance. 
By contrast, the throughput of \textit{PostMHL} rises dramatically for SC with a longer $\delta t$ due to the engagements of faster query processing.
Figure~\ref{fig:varyParameter} (g)-(i) show the result of varying response time $R^*_q$. Similar to $\delta t$, the throughput of most baselines remains steady under larger $R^*_q$, while PostMHL generally has a significant throughput improvement.}
Overall, \textit{PostMHL} generally outperforms baseline and exhibits remarkable throughput improvement in most cases.


\colorR{\textbf{Exp 6: Effect of Thread Number $p$.}}
\colorR{
Figure~\ref{fig:threadNum} presents the performance speedup when varying $p$ from $1$ to $160$, where each legend is a combination of an algorithm (\textit{PMHL} or \textit{PostMHL}) and a dataset (NY or FLA). The update speedup increases with the number of threads but reaches a plateau due to the unparalleled overlay index update and limited partition number. \textit{PostMHL} has a higher speedup ($9.3\times$) than \textit{PMHL} ($2.3\times$) since its partition number ($k_e=32$) is larger than \textit{PMHL}'s ($k=8$). As shown in Figure~\ref{fig:threadNum}-(b), the tendency of throughput speedup is similar to update time: rise dramatically first and then reach a plateau. In particular, the throughput of \textit{PMHL} has up to $280\times$ speed up. This is because the initial throughput of \textit{PMHL} (when $p=1$) is quite low while a faster update time powered by more threads could vacate more time for efficient cross-boundary query processing.}

\section{Related Work}
\label{sec:Related}
We discuss related works by classifying them into two types:

\colorR{\underline{\textit{Shortest Path Algorithms}}}.
The index-free algorithms like \textit{Dijkstra's}~\cite{dijkstra1959note}, $A^*$~\cite{Astar_hart1968formal}, and \textit{BiDijsktra}~\cite{bidijk_nicholson1966finding} search the graph on the fly so they are naturally dynamic~\cite{thomsen2012effective,li2020fast,zhang2020stream} but they are slow for query. 
Index-based methods were proposed \cite{highway_sanders2005highway,CH_geisberger2008contraction,2hop_cohen2003reachability,HL_abraham2011hub,HHL_abraham2012hierarchical,PLL_akiba2013fast,PSL_li2019scaling,CoreTree_li2020scaling,H2H_ouyang2018hierarchy,P2H_chen2021p2h} to achieve 1-5 orders of magnitude of improvement, with \textit{H2H}~\cite{H2H_ouyang2018hierarchy} and \textit{P2H}~\cite{P2H_chen2021p2h} performing best on road networks.
To adapt to dynamic scenarios, index maintenance~\cite{IncPLL_akiba2014dynamic,DecPLL_d2019fully,IUPLL_qin2017efficient,DPSL_zhang2021efficient,DPSL_zhang2023parallel,DCH_geisberger2012exact,DCH_wei2020architecture,DCH_ouyang2020efficient,DH2H_zhang2021dynamic,BHL_farhan2022batchhl,EPSP_zhang2023universal,DCT_zhou2024scalable} has also been studied. 
Among them, \textit{DCH}~\cite{DCH_ouyang2020efficient} and \textit{DH2H}~\cite{DH2H_zhang2021dynamic} are two state-of-the-art methods.
It is worth mentioning that \textit{DCH}~\cite{DCH_ouyang2020efficient} outperforms popular industrial solution \textit{CRP}~\cite{CRP_delling2017customizable} in update time. 
\colorR{
However, most of these methods only focus on reducing query time or update time, ignoring that high throughput on dynamic networks requires both high query and update efficiency. 
It is worth noting that the throughput of kNN queries has also been studied in previous works~\cite{TOAIN_luo2018toain,GLAD_he2019efficient}. \textit{TOAIN}~\cite{TOAIN_luo2018toain} is an adaptive algorithm for optimizing the throughput of $k$NN queries based on a multi-level CH index called \textit{SCOB}. However, they were designed for static networks and using them for random point-to-point SP calculations is wasteful.}

\colorR{\underline{\textit{Partitioned Shortest Path Index}}.
To scale up to large graphs, the PSP index~\cite{Gtree_zhong2013g,GstarTree_li2019g,CRP_delling2017customizable,CCH_dibbelt2016customizable,ROAD_lee2010road,CoreTree_li2020scaling,li2019time,liu2021efficient,liu2022fhl,wang2016effective,li2020fastest,liu2022multi} has also been investigated and can be classified into three types based on partition structure~\cite{EPSP_zhang2023universal}: 
the \textit{Planar PSP}~\cite{TNR_bast2007transit,ArcFlag_hilger2009fast,CRP_delling2017customizable} treats all partitions equally on one level, such as \textit{TNR}~\cite{TNR_bast2007transit}, \textit{PCH}~\cite{EPSP_zhang2023universal} and \textit{PH2H}~\cite{EPSP_zhang2023universal}. 
The \textit{Hierarchical PSP} such as \textit{CRP}~\cite{CRP_delling2017customizable}, \textit{G-Tree}~\cite{Gtree_zhong2013g} and \textit{ROAD}~\cite{ROAD_lee2010road} organizes the partitions hierarchically and each level is a planar partition.
\textit{Core-Periphery PSP}~\cite{CoreTree_li2020scaling,DCT_zhou2024scalable,sketch_wang2021query} treats the partitions discriminately by taking some important vertices as “Core” and the remaining ones as “Peripheries”, which is more suitable for small-world networks. 
To adapt them to dynamic networks, \cite{EPSP_zhang2023universal} puts forward three general PSP strategies and a universal scheme for designing the dynamic PSP index. Nevertheless, these dynamic PSP indexes sacrifice the query efficiency, thus failing to achieve high throughput query processing.
}
\section{Conclusion}
\label{sec:Conclusion}
In this paper, we investigate high throughput query processing on dynamic road networks.
We first propose a \textit{cross-boundary PSP strategy} with an insightful analysis of the \textit{\colorR{upper bound} of PSP index query efficiency}.
Then we propose the \textit{PMHL} index to leverage thread parallelization and multiple PSP strategies for fast index update and continuously improved query efficiency. Lastly, we propose \textit{PostMHL} by adopting \textit{TD partitioning} to further enhance throughput by faster query and index update efficiency.
The experiments demonstrate that \textit{PostMHL} can yield up to 2 orders of magnitude throughput improvement compared to the state-of-the-art baselines, making hop-based solutions more practical for real-life applications. 

\section*{Acknowledgments}
This work was supported by Natural Science Foundation of China \#62202116, Hong Kong Research Grants Council grant\# 16202722,  Guangzhou‑HKUST(GZ) Joint Funding Scheme \#2023A03J0135, Guangzhou Basic and Applied Basic Research Scheme \#2024A04J4455, Guangdong-Hong Kong Technology Innovation Joint Funding Scheme \#2024A0505040012, Guangzhou HKUST Fok Ying Tung Research Institute grant \#2023ZD007, Guangzhou municipality big data intelligence key lab \#2023A03J0012, and JC STEM DSF Lab funded by Hong Kong Jockey Club Charities Trust.
\balance



\bibliographystyle{IEEEtran}
\bibliography{Reference}

\newpage
\newpage

\section{Appendix}\label{sec:appendix}

\subsection{Index Update of No-boundary and Post-boundary Strategy}
As introduced in Section~\ref{subsec:Preliminary_PSPIndex}, the index update of no-boundary and post-boundary strategies are similar to their index construction by replacing the build with the update. Nevertheless, we introduce their update procedures as follows.

\begin{figure}[h]
	\centering
	\includegraphics[width=1\linewidth]{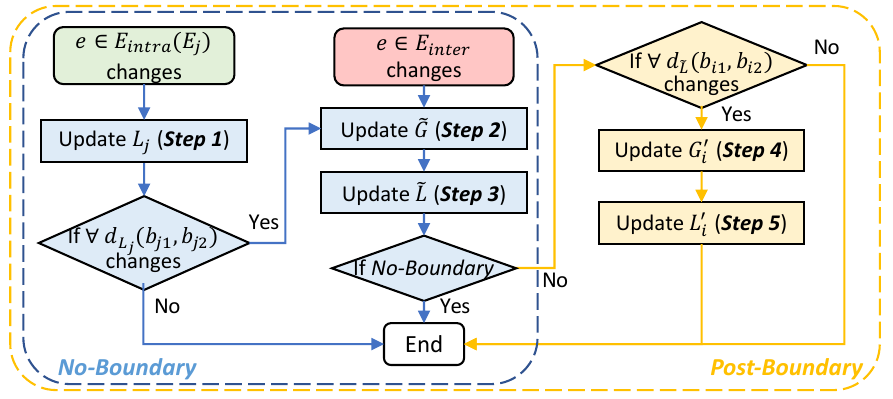}
	\caption{Index Update of No-boundary and Post-boundary Strategy}\label{fig:NoBoundaryUpdate}
\end{figure}

\textbf{Index Update for No-boundary Strategy.}
Since the weight change of inter-edges does not affect the index for each subgraph, we divide index updates into two scenarios as shown in Figure~\ref{fig:NoBoundaryUpdate}. Scenario 1: Inter-edge weight change. When  $e\in E_{inter}$ changes, only $\tilde{L}$ needs an update; Scenario 2: Intra-edge weight change. When $e\in E_{intra}$ ($e\in E_j$) changes, we first update $L_j$ and compare the old and new weights of $e(b_{j1},b_{j2})$ between boundary in $G_j$. If there is an edge weight update, we need to further update $\tilde{L}$. 

\textbf{Index Update for Post-boundary Strategy.}
It is similar to \textit{No-Boundary} with an additional judgment and processing shown in Figure~\ref{fig:NoBoundaryUpdate}. Scenario 1: Intra-edge weight change. Suppose $e\in E_i$ changes, we update $G_i, L_i$ and then update $\tilde{G}, \tilde{L}$ if any $d_{L_j}(b_{j1},b_{j2})$ changes. Then we update $\{G'_i\}, \{L'_i\}$ if $d_{G_i'}(d_{i1},d_{i2})$ and $d_{\tilde{L}}(d_{i1},d_{i2})$ are different; Scenario 2: Inter-edge weight change. Suppose $e\in E_{inter}$ changes, we update $\tilde{G}, \tilde{L}$ and then update $\{G_i'\}, \{L_i'\}$.

\subsection{Pseudo-code of PMHL Index Construction}
The pseudo-code of \textit{PMHL} index construction is presented in Algorithm~\ref{algo:PMHLConstruct}. \textit{PMHL} first leverages graph partitioning method \textit{PUNCH}~\cite{PUNCH_delling2011graph} to divide the original road network $G$ into $k$ subgraphs (Line~\ref{pmhl:1}). Given the partition results as input, a \textit{boundary-first} vertex ordering method~\cite{liu2022fhl} is leveraged to obtain the vertex order $r$ of all vertices, which assigns the boundary vertices higher ranks than the non-boundary vertices since the \textit{cross-boundary strategy} should fulfill the boundary-first property (Line~\ref{pmhl:2}).
After obtaining the vertex order, there are six main steps for constructing the \textit{PMHL} index (Lines 3-\ref{pmhl:10}), obtaining the index of various PSP strategies (see Section~\ref{subsec:PMHLIndexing} for details).

\begin{algorithm}[b]
	\caption{PMHL Index Construction}
	\label{algo:PMHLConstruct}
    \scriptsize
	\LinesNumbered
    \KwIn{Road network $G=\{V,E\}$}
	\KwOut{PMHL index $L_{PMHL}=\{\tilde{L},\{L_i\},\{L'_i\},L^{*}\}$}
    \SetKwBlock{DoParallel}{parallel\_for}{end}
    $\{G_i|1\leq i\leq k\}\gets $ Partitioning $G$ by PUNCH~\cite{PUNCH_delling2011graph};\Comment{Get partition graphs}\label{pmhl:1} \\
    $r\gets$ \textsc{BoundaryFirstOrder($\{G_i\}$)};\Comment{Boundary-first vertex ordering}\label{pmhl:2} \\
    // No-boundary Index Construction\\
    \DoParallel($i \in [1,k]$)
    {
        $L_i\gets$ \textsc{MHLIndexing($G_i$)}; \Comment{Step 1: Build partition indexes $\{L_i\}$}\label{pmhl:4}\\
    }
    $\tilde{G}\gets$ \textsc{OverlayGraphBuild($\{L_i\}$)}; \Comment{Step 2: Build overlay graph}\label{pmhl:5}\\
    $\tilde{L}\gets$ \textsc{MHLIndexing($\tilde{G}$)}; \Comment{Step 3: Build overlay index $\tilde{L}$}\label{pmhl:6}\\
    // Post-boundary Index Construction\\
    \DoParallel($i \in [1,k]$)
    {
    $G'_i\gets$ \textsc{GetExtendedGraph($\tilde{L},G_i$)}; \Comment{Step 4: Build extended partitions}\label{pmhl:8}\\
    $L'_i\gets$ \textsc{MHLIndexing($G'_i$)}; \Comment{Step 5: Build post-boundary index $\{L'_i\}$}\label{pmhl:9}\\
    }
    // Cross-boundary Index Construction, see Algorithm~\ref{algo:TDAggregation}\\
    $L^{*}\gets$\textsc{CrossIndexBuild($\tilde{L},\{L_i\}$)}; \Comment{Step 6: Build cross-boundary index $L^{*}$}\label{pmhl:10}\\
    \Return $L_{PMHL}=\{\tilde{L},\{L_i\},\{L'_i\},L^{*}\}$;\\
\end{algorithm}

\subsection{Pseudo-code of PostMHL Index Construction}
The pseudo-code of \textit{PostMHL} index construction is presented in Algorithm~\ref{algo:PostMHLConstruct}. 
We first utilize \textit{MDE-based tree decomposition}~\cite{MDETD_xu2005tree} to obtain $T$ and shortcut arrays for all tree nodes (Line~\ref{pomhl:1}). 
After that, \textit{TD-partitioning} is employed to obtain the graph partition results, and the overlay index is built in a top-down manner like the \textit{H2H} index (Line~\ref{pomhl:3}). 
After getting the overlay index, we build the post-boundary index for all partitions in parallel (Lines 5-\ref{pomhl:31}). 
In particular, given the partition $G_i$, we first compute the all-pair distances among the boundary vertices of $G_i$ based on the overlay index and then store the results in a map table $D$ for later reference (Lines \ref{pomhl:6}-\ref{pomhl:8}). 
Sourcing from the root vertex, the post-boundary index, including the boundary array and distance array entries of in-partition ancestors, is conducted in a top-down manner according to the \textit{minimum distance property}  \cite{DH2H_zhang2021dynamic}, \ie $d(v,b)=\min_{u\in X(v).N}\{sc(v,u)+d(u,b)\},\forall v\in G_i, b\in B_i\cup X(v).A$ 
(Lines 9-\ref{pomhl:31}). 
Lastly, the cross-boundary index is constructed parallelly in a top-down manner from each partition's root vertex just like the overlay index construction (Lines 32-\ref{pomhl:33}).

\subsection{Proofs}

\noindent\textbf{Proof of Theorem~\ref{theorem:CrossBoundaryCorrect}.}
We first prove the distance array of $L^*$ is correct in two cases:
    
Case 1: $v\in \tilde{G}$. This is naturally correct as the cross-boundary tree inherits the node relationship of the overlay tree and leverages the neighbor set of $v$'s overlay index as the vertex separator.

Case 2: $v\notin \tilde{G}, v\in G_i$. As per Case 1, the cross-boundary labels of all $G_i$'s boundary vertices are accurate. 
Therefore, the distances from $v$ to its ancestor $u\notin G_i$ computed by the top-down label construction are correct as it essentially uses $B_i$ as the vertex separator. 
Meanwhile, the distances from $v$ to its ancestor $u\in G_i$ are also accurate, as it uses the neighbor set of $v$'s partition index as the vertex separator.

We next prove $\forall s\in G_i, t\in G_j, i\neq j$, the LCA of $X^{*}(s)$ and $X^{*}(t)$ is the vertex separator of $s$ and $t$ in four cases:
	
Case 1: $s\in \tilde{G}, t\in \tilde{G}$. The LCA of $X^{*}(s)$ and $X^{*}(t)$ (denoted as $LCA(X^{*}(s),X^{*}(t))$) is the same as $LCA(\tilde{X}(s),\tilde{X}(t))$ since $L^{*}$ has equivalent node relationships among the overlay vertices of $\tilde{L}$. Hence, $LCA(X^{*}(s),X^{*}(t))$ is the vertex separator.

Case 2: $s\in \tilde{G}, t\notin \tilde{G}$. We take the concise form of $sp(s,t)$ by extracting only the boundary vertices as $sp_c=\left<s=b_0,\dots,b_n,t\right>$ ($b_i\in B, 0\le i\le n$). As per Case 1, we have $LCA(X^{*}(s),X^{*}(b_n))$ is the vertex separator of $s$ and $b_n$. Besides, $X^{*}(b_n)$ is the ancestor of $X^{*}(t)$ since $L^{*}$ inherits the subordinate relationships of the partition trees and $r(b_n)>r(t)$. Therefore, $LCA(X^{*}(s),X^{*}(t))$ is the same as $LCA(X^{*}(s),X^{*}(b_n))$ and it is the vertex separator.

Case 3: $s\notin \tilde{G}, t\in \tilde{G}$. The proof is similar to Case 2.

Case 4: $s\notin \tilde{G}, t\notin \tilde{G}$. We take the concise form of $sp(s,t)$ by extracting only the boundary vertices as $sp_c=\left<s,b_0,\dots,b_n,t\right>$ ($b_i\in B, 0\le i\le n$). As per Case 1, we have $LCA(X^{*}(b_0),X^{*}(b_n))$ is the vertex separator of $b_0$ and $b_n$. Besides, $X^{*}(b_0)$ is the ancestor of $X^{*}(s)$ while $X^{*}(b_n)$ is the ancestor of $X^{*}(t)$. Therefore, $LCA(X^{*}(s),X^{*}(t))$ is the same as $LCA(X^{*}(b_0),X^{*}(b_n))$ and it is the vertex separator.

Since the distance array is correct and the LCA is a vertex separator, $L^*$ can correctly answer cross-partition queries.
\hfill $\square$\par\vspace{0.5em}

\begin{algorithm}[t]
	\caption{PostMHL Index Construction}
	\label{algo:PostMHLConstruct}
    \scriptsize
	\LinesNumbered
    \KwIn{Road network $G$, bandwidth $\tau$, expected partition number $k_e$, lower and upper bound of partition size $\beta_l, \beta_u$}
	\KwOut{PostMHL index $L=\{X(v)|v\in V\}$}
    \SetKwBlock{DoParallel}{parallel\_for}{end}
    $T\gets$ \textsc{TreeDecomposition($G$)};\Comment{MDE-based tree decomposition}\label{pomhl:1}\\ 
    $\tilde{G},\{G_i|i\in[1,k]\}\gets$ \textsc{TDPartition($T,\tau,k_e,\beta_l,\beta_u$)};\Comment{TD-partitioning}\label{pomhl:2}\\
    \textsc{OverlayIndexing}($\tilde{G}$,$L$); \Comment{Top-down overlay index construction}\label{pomhl:3}\\
    // Post-boundary index construction\label{pomhl:4}\\
    \DoParallel($i\in [1,k]$)
    {
        $u\gets$ root vertex of $G_i$;\label{pomhl:6}\\
        \For{$b_1\in X(u).N$ and $b_2\in X(u).N$}
        {
            $D[b_1][b_2]\gets Q(b_1,b_2)$; \Comment{Get all-pair boundary distances}\label{pomhl:8}\\
        }
        \For{$v\in G_i$ and $X(v)\in T$ in a top-down manner}
        {
            Suppose $X(v).N=(x_1,x_2,...)$;\label{pomhl:10}\\
            \For{$j=1$ to $|X(v).N|$}
            {
                $X(v).pos[j]\gets $the position of $x_j$ in $X(v).A$;\label{pomhl:12}
            }
            // Compute boundary array\label{pomhl:13}\\
            \For{$j=1$ to $|X(u).N|$}
            {
                $X(v).disB[j]\gets \infty$;\label{pomhl:15}\\
                \For{$k=1$ to $|X(v).N|$}
                {
                   \textbf{if} $x_k\in \tilde{G}$ \textbf{then}
                    $d\gets X(v).sc[k]+D[x_j][x_k]$;\label{pomhl:17}\\
                   \textbf{else} $d\gets X(v).sc[k]+X(x_k).disB[j]$;\label{pomhl:18}\\
                   $X(v).disB[j]\gets\min\{X(v).disB[j],d\}$;\label{pomhl:19}\\
                }
            }
            // Compute distance array\label{pomhl:20}\\
            \For{$j=1$ to $|X(v).A|-1$}
            {
                \textbf{if} $X(v).A[j]\in \tilde{G}$ \textbf{then} continue; \Comment{Prune overlay ancestors}\label{pomhl:22}\\
                $c\gets X(v).A[j]; X(v).dis[j]\gets \infty$;\label{pomhl:23}\\
                \For{$k=1$ to $|X(v).N|$}
                {
                    \If{$x_k\in \tilde{G}$}
                    {
                        $l\gets$ position of $x_k$ in $X(u).N$; $d\gets X(c).disB[l]$;\label{pomhl:26}\\
                    }
                    \Else{
                        \textbf{if} $X(v).pos[k]>j$ \textbf{then} $d\gets X(x_k).dis[j]$;\label{pomhl:28}\\
                        \textbf{else} $d\gets X(c).dis[X(v).pos[k]]$;\label{pomhl:29}\\
                    }
                    $X(v).dis[j]\gets\min\{X(v).dis[j],X(v).sc[k]+d\}$;\label{pomhl:30}\\
                }
            }
            $X(v).dis[|X(v).A|]\gets 0$;\label{pomhl:31}\\
        }
    }
    \DoParallel($i\in [1,k]$)
    {
        \textsc{CrossPartiIndexing}($G_i,L$);\Comment{Cross-boundary index construction}\label{pomhl:33}\\
    }
    \Return $L$;\\
    
\end{algorithm} 

\noindent\textbf{Proof of Theorem~\ref{theorem:overlayIndex}.}
The post-boundary index needs the overlay index to construct all-pair boundary shortcuts and check whether any boundary shortcut has changed when updating it. On the other hand, the cross-boundary index requires the overlay index for index construction and to identify the affected in-partition vertices during index maintenance. Since they are independent of each other and only rely on overlay index, the theorem is proved.
\hfill $\square$\par\vspace{0.5em}

\noindent\textbf{Proof of Theorem~\ref{theorem:PostMHLComplexity}.}
    The index size of \textit{PostMHL} equals the H2H index size $O(n\cdot h)$~\cite{H2H_ouyang2018hierarchy} plus boundary array size $O(n_p\cdot |B_{max}|)$, where $|B_{max}|=\max_{i\in[1,k]}\{|B_i|\}$ is maximal boundary vertex number of all partitions, $n_p$ is the in-partition vertex number. Hence, index size is $O(n\cdot h+n_p\cdot |B_{max}|)$. 
    The indexing time of \textit{PostMHL} consists of tree decomposition $O(n\cdot(w^2+\log(n)))$, TD-partitioning $O(n\log(n))$, overlay indexing $O(n_o\cdot(\tilde{h}\cdot \tilde{w}))$, thread-parallel post-boundary and cross-boundary indexing $O(\max_{i\in[1,k]}\{n_i\cdot(h\cdot w+|B_i|\cdot w)\})$. 
    As for index update, according to \cite{DH2H_zhang2021dynamic}, the H2H decrease update and increase update complexity is $O(w\cdot(\delta+\Delta h))$ and $O(w\cdot(\delta+\Delta h\cdot(w+\epsilon)))$, respectively. Therefore, the post-boundary and cross-boundary index updates can be conducted in parallel for all partitions.
\hfill $\square$\par\vspace{0.5em}

\subsection{Additional Experimental Results}

\begin{figure}[h!]
	\centering
	\includegraphics[width=1\linewidth]{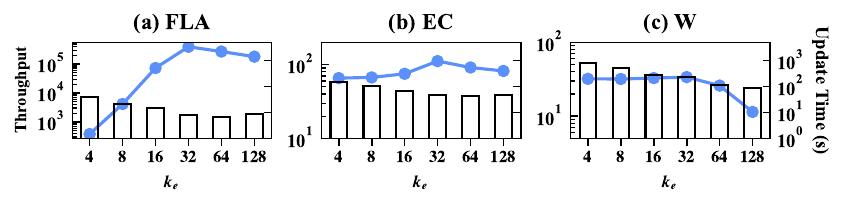}
	\caption{Effect of $k_e$ for PostMHL (Polyline: $\lambda^*_q$, Bar: $t_u$)}\label{fig:partiNumPost}
\end{figure}

\textbf{Exp 7: Effect of Partition Number $k_e$ for PostMHL.}
We evaluate the effect of partition number $k_e$  for \textit{PostMHL}. 
Figure~\ref{fig:partiNumPost} presents the throughput $\lambda^*_q$ and update time $t_u$ on datasets FLA, EC, and W.
Similar to PMHL, both small and large $k_e$ result in reduced throughput. In general, $k_e=32$ performs the best due to the balance between update workload and parallelization. A larger $k_e$ can better exploit parallelization but increase the overlay vertex number due to smaller partition size, deteriorating the update efficiency of overlay and cross-boundary index. 

\begin{figure}[h!]
	\centering
	\includegraphics[width=1\linewidth]{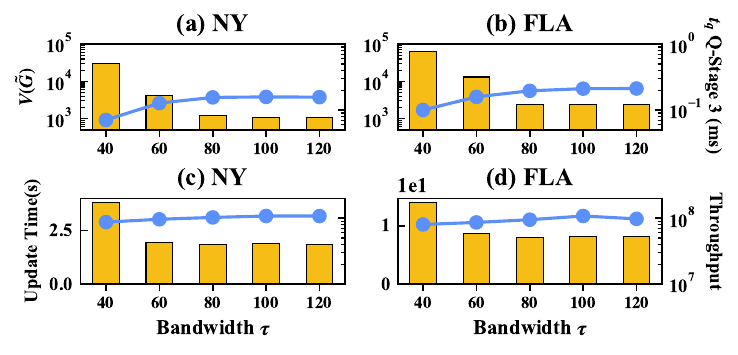}
	\caption{Effect of $\tau$ for PostMHL (Bar: left axis, Polyline: right axis)}\label{fig:bandwidthQT}
\end{figure}

\textbf{Exp 8: Effect of Bandwidth for PostMHL.}
We further demonstrate the effect of bandwidth $\tau$ \textit{PostMHL}. 
Figure~\ref{fig:bandwidthQT} presents the results on datasets NY and FLA.
In particular, Figure~\ref{fig:bandwidthQT} (a)-(b) shows the overlay vertex number and query time of Q-Stage 3. 
Observe that the increase in bandwidth leads to smaller overlay vertex numbers since a large $\tau$ leads to more vertices becoming the partition vertex. 
However, a large $\tau$ deteriorates the query efficiency of Q-Stage 3 (post-boundary query). This is due to two reasons. Firstly, the proportion of cross-boundary queries increases as the overlay graph size reduces. Secondly, the cross-boundary query efficiency is undermined due to the larger boundary vertex number (\ie bandwidth). It is worth noting that the change of bandwidth could not affect the query efficiency of Q-Stage 1, 2, and 4 of PostMHL since the graph that \textit{BiDijkstra} and \textit{PCH} search on and the \textit{cross-boundary index} do not change.
Figure~\ref{fig:bandwidthQT} (c)-(d) depicts the update time and query throughput.
We can see that a small $\tau$ tends to deteriorate the update time and query throughput. This is because a small bandwidth leads to a larger overlay graph, whose index maintenance cannot be accelerated by multi-thread parallelization, thus reducing the index update efficiency and throughput. Therefore, it is generally advisable to set a large bandwidth for PostMHL even though a large $\tau$ could undermine the post-boundary query processing. 

\end{document}